%% file: main.tex
\documentclass[twocolumn,longbibliography]{revtex4-2} 
\usepackage{amsmath,amsbsy,dsfont}
\usepackage{amssymb}
\usepackage{hyperref}
\usepackage{tikz}
\usepackage{subfigure}
\usetikzlibrary{arrows, decorations.markings}
\usepackage{rotating}
\usepackage{xcolor}
\usepackage{graphicx}

\usepackage[version=4]{mhchem}
\usepackage{cleveref}
\usepackage[]{units}

\usepackage{booktabs}      
\usepackage{multirow}

\usepackage{enumitem}
\usepackage[normalem]{ulem}

\newcommand{\bs}[1]{\boldsymbol{#1}} 
\renewcommand{\vec}{\bs}
\newcommand{\mub}{\,\mu_\text{B}}
\newcommand{\meV}{\,\text{meV}}
\newcommand{\muth}{\mu_\text{th}}
\newcommand{\muexp}{\mu_\text{exp}}

\input{data.tex}

\begin{document}
\title{Benchmark for \textit{ab initio} prediction of magnetic~structures based on cluster~multipole~theory}
\author{M.-T. Huebsch$^{1,2}$, T. Nomoto$^2$, M.-T. Suzuki$^{3,4}$ and R. Arita$^{1,2}$}
\affiliation{$^1$Center for Emergent Matter Science, RIKEN, Wako, Saitama 351-0198, Japan\\
$^2$University of Tokyo, 7-3-1 Hongo, Bunkyo-ku, Tokyo 113-8656, Japan\\
$^3$Center for Computational Material Science, Institute for Materials Research, Tohoku University, Sendai, Miyagi 980-8577, Japan\\
$^4$Center for Spintronics Research Network, Graduate School of Engineering Science, Osaka University, Toyonaka, Osaka 560-8531, Japan}
\date{\today}

\begin{abstract}
  The cluster multipole (CMP) expansion for magnetic structures provides a scheme to systematically generate  candidate magnetic structures specifically including noncollinear magnetic configurations adapted to the crystal symmetry of a given material. A comparison with the experimental data collected on MAGNDATA shows that the most stable magnetic configurations in nature are linear combinations of only few CMPs. Furthermore, a high-throughput calculation for all candidate magnetic structures is performed in the framework of spin-density functional theory (SDFT). We benchmark the predictive power of CMP+SDFT with $\totcandidates$ calculations, which show that 
  (i) the CMP expansion administers an exhaustive list of candidate magnetic structures, 
  (ii) CMP+SDFT can narrow down the possible magnetic configurations to a handful of computed configurations, and
  (iii) SDFT reproduces the experimental magnetic configurations with an accuracy of $\pm0.5\mub$.
  For a subset the impact of on-site Coulomb repulsion $U$ is investigated by means of 1545 CMP+SDFT+U calculations revealing no further improvement on the predictive power.
\end{abstract}

\maketitle
\section{Introduction} \label{sec:Introduction}

The grand challenge in first-principles calculation for magnetic materials is whether we can predict the experimental magnetic structure for a given material.
Among a variety of possible functional materials, noncollinear magnets are a fascinating playground for materials design \cite{schleder2019dft,zhang2020high} as they facilitate a wide range of fundamental phenomena and possible applications.

For example, in the context of antiferromagnetic (AFM) spintronics~\cite{Baltz2018} there is a particular interest in noncollinear antiferromagnetism
sparked by
(i) its robustness against perturbations due to magnetic fields,
(ii) a quasi-absence of magnetic stray fields disturbing for instance nearby electronic devices, and
(iii) ultrafast dynamics of AFM domainwalls~\cite{Nomoto2020}, as well as
(iv) its ability to generate large magnetotransport effects~\cite{kleiner1966space,kleiner1967space,Seemann2015}.
Hence, the optimization of AFM materials would open the door for applications such as seamless and low-maintenance energy generation, ultrafast spintronics and robust data retention, as well as be a guide towards advancing fundamental understanding of magnetotransport.

However, first-principles calculations with the generalized gradient approximation (GGA) in the framework of spin-density functional theory (SDFT) for magnetic materials have a problem: It is still an open question how accurately SDFT--GGA can reproduce the experimental magnetic ground state. While SDFT has been widely used in studies on various magnets~\cite{Kuebler2017}, there has been no systematic benchmark calculation for noncollinear AFM materials.
Previous attempts have been restricted to collinear magnetism~\cite{Horton2019} or even stricter symmetry constrains~\cite{sanvito2017accelerated,stevanovic2012correcting,gorai2016thermoelectricity}.
In regard to noncollinear AFM materials, high-throughput calculations have been limited to setting the experimentally determined magnetic configuration as an initial guess~\cite{Xu2020}.
A recently proposed attempt to predict magnetic structures based on a genetic evolution algorithm \cite{zheng2020maggene} strongly relies on the proper prediction of the magnetic ground state by SDFT.
The lack of a systematic benchmark calculation is a consequence of the fact that it is a highly non-trivial task to investigate all the local minima in the SDFT energy landscape. Indeed, to search for all the (meta-)stable states, we need an exhaustive list of physically reasonable magnetic configurations for which first-principles calculations can be performed.

To this end, we devise the so-called cluster multipole (CMP) expansion, which enables the expansion of an arbitrary magnetic configuration in terms 
of an orthogonal basis set of magnetic multipole configurations.
By means of the CMP expansion, a list of initial magnetic structures for self-consistent GGA calculations is efficiently and systematically generated. 
With this at hand, a systematic high-throughput calculation with $\totcandidates$ calculations has been performed.

The structure of the paper is as follows: In~\Cref{Sec:Method}, we explain the basic idea of CMP (\Cref{Subsec:Cluster Multipole expansion}) and setup for GGA calculation (\Cref{Subsec:Setup for SDFT}). In~\Cref{Sec:Results}, the magnetic configuration of $\totmat$ materials is predicted using a combination of the CMP expansion and SDFT (CMP+SDFT). 
A comparison to the experimental data shows that the magnetic ground state can be narrowed down to be among a handful of computed configurations
and SDFT reproduces the experimental on-site magnetic moment with an accuracy of approximately $\pm0.5\mub$.
This benchmark, which is summarized in~\Cref{Sec:Conclusion and Outlook}, thus provides a solid foundation for the 
\textit{ab initio} predictions of various magnetic properties.

\section{Methods} \label{Sec:Method}

In this section we shortly discuss the employed methods, namely the CMP expansion and SDFT. 
As the CMP expansion is a rather novel approach~\cite{suzuki2017cluster,suzuki2018first,huyen2019topology}, it shall be motivated and set out in some detail. 
However for more background and details of the algorithm we refer the reader to Ref.~\cite{Suzuki2019}. 
SDFT on the other hand is a well established method~\cite{Barth1972,Kuebler2017}. 
It is available as part of many \textit{ab initio} packages~\cite{Hobbs2001,Corso2005,elk,Eich2013,exitingcode,Sharma2007} in its generalized version~\cite{Nordstrom1996,Eschrig1999}, which is applicable to noncollinear AFM configurations.
Here, we chose to use VASP~\cite{Kresse1993,Kresse1994,Hobbs2001} and hence we merely elaborate on the 
setup details employed in this study.

\subsection{Cluster Multipole expansion} \label{Subsec:Cluster Multipole expansion}

The cluster multipole (CMP) expansion for magnetic structures~\cite{Suzuki2019,suzuki2017cluster} provides an orthogonal basis set of magnetic configurations, which are symmetrized based on the crystallographic point group.
In order to motivate the expansion, let us consider the vector Poisson equation:
\begin{equation}
   \nabla^2 \vec{A}(\vec{r}) = - \frac{4\pi}{c} \vec{j}(\vec{r}),
\end{equation}
where $\vec{j}(\vec{r})=c\nabla \times \vec{M}(\vec{r})$ is the current density and $\vec{M}(\vec{r})$ is the magnetization density.
Here, the Coulomb gauge $\nabla \cdot \vec{A}(\vec{r}) = 0$ is invoked and the potential outside of the magnetization density is considered.
The rotational invariance of $\nabla^2$ allows the vector gauge potential $\vec{A}(\vec{r})$ to be 
expanded w.r.t.\ vector spherical harmonics $\vec{Y}^{l1}_{pq}$ \cite{Kusunose2008}.
Accordingly, the magnetic field $\vec{B}(\vec{r})=\vec{\nabla}\times \vec{A}(\vec{r})$ can be written in terms of magnetic multipole moments $M_{lm}$ as follows \cite{hayami2018classification}
\begin{align}
    \vec{B}(\vec{r})=- \sum_{l=1}^{\infty} \sum_{m=-l}^l \sqrt{4\pi(l+1)} M_{lm} \frac{1}{r^{l+2}} \vec{Y}^{l+1,1}_{lm}(\Omega),
\end{align}
where $l$ is the orbital angular momentum quantum number and $m$ magnetic quantum number.


Following M.-T.\ Suzuki {\it et al.\ }in Ref.\ \cite{Suzuki2019} the CMPs for a magnetic configuration on a point form
$\left| m \right> = \left(\vec{m}_1, \vec{m}_2,...,\vec{m}_{N} \right)^T$ read
\begin{equation}
  M_{lm} = \sqrt{\frac{4\pi}{2l+1}} \sum_{i=1}^{N} \vec{m}_i \cdot \left[ \nabla \left( \left| \vec{r}_i\right|^l Y^{l*}_{m} \right) \right].
\end{equation}
$\vec{m}_i$  is a local magnetic moment on the magnetic site $i$ at position $\vec{r}_i$.
For a given point group the point form is a set of all symmetrically equivalent points and can be classified into Wyckoff positions~\cite{Hahn2006} in analogy to the Wyckoff positions of space groups.
Here, $N$ is the multiplicity of the Wyckoff position of the point form, that constitutes the magnetic configuration. As introduced by Ref.\ \cite{Suzuki2019} a point form carrying a magnetic configuration is referred to as (magnetic) cluster in the context of the CMP expansion for magnetic structures. In contrast to Ref.\ \cite{Suzuki2019}, here we do not introduce toroidal moments.

Symmetrization according to irreducible representations of the crystallographic point group allows for a physically meaningful expansion
w.r.t.\ symmetrized harmonics 
\begin{equation}
  {\mathcal Y}_{l\gamma} = \sum_{m} c^\gamma_{lm} Y_{lm}, 
\end{equation}
where $\gamma$ indicates the irreducible representation including the existing components of it.
Here, the tabulated coefficients~\cite{Kusunose2008} $c^\gamma_{lm}$ are chosen to be real valued.
With this a \textit{virtual cluster}~\cite{Suzuki2019} is constructed, where each magnetic site is assigned a magnetic moment.
By mapping $l\gamma \to n$ through a Gram-Schmidt orthonormalization scheme the CMP basis is computed. 

The CMP basis can be written as
\begin{equation}
  \left\{ \left| n \right> = \left(\vec{e}^{(n)}_1, \vec{e}^{(n)}_2,...,\vec{e}^{(n)}_{N} \right)^T \right\},
  \label{eq:cmp basis}
\end{equation}
where $\vec{e}^{(n)}_i$ is a unit vector of a local magnetic moment on the magnetic site $i$.
By convention $n=1,2,3$ corresponds to ferromagnetism, while  $n \ge 4$ corresponds to more complicated higher order magnetic configurations including noncollinear magnetism.
The definition of $\left|n\right>$  coincides with $\left\{\vec{e}^\mu_{l\gamma}\right\}$ in Ref.~\cite{Suzuki2019}
up to the choice of normalization~\footnote{ In other words,
${\bf e}^{(n)}_i$ correspond to components of $\left\{\vec{e}^\mu_{l\gamma}\right\}$ in Ref.~\cite{Suzuki2019} except that here $\sum_i^N {\bf e}^{(n)}_i {\bf e}^{(n')}_i = N\,\delta_{nn'}$, while in Ref.~\cite{Suzuki2019}
$\left(\left\{\vec{e}^n\right\}\cdot\left\{\vec{e}^{n'}\right\}\right)=\delta_{nn'}$. 
Also note that $\left\{\vec{e}^\mu_{l\gamma}\right\}$ is labeled by $\mu=1,2$ representing magnetic (M) and magnetic toroidal (MT) multipoles, respectively.
As we do not expand in terms of MT multipoles, those components appear as higher order magnetic multipoles here, and thus completeness is still ensured.}.

In case that the period of the magnetic order coincides with that of the crystal structure, the propagation vector of the magnetic order $\vec{q}$ is zero. 
The magnetic structure is said to exhibit $\vec{q}=\vec{0}$ magnetism.
Note that $3$ continuous degrees of freedom of rotation of the magnetic moment per magnetic site for a total of $N$ magnetic sites yields $3N$ linearly independent 
magnetic configurations and thus $n=1,...,3N$. 
In this work, the configuration space of $\vec{q}=\vec{0}$ magnetic structures is explored.

The CMP basis defined in \Cref{eq:cmp basis} is complete
\begin{equation}
	\frac{1}{N} \sum_{n=1}^{3N}\left|n\right>\left<n\right| = \mathds 1_{3N\times3N},
\end{equation}
and obeys the orthogonality relation
\begin{equation}
	\left<n|n'\right> = N\,\delta_{nn'}.
\end{equation}
Finally, the symmetry-adapted CMP coefficient reads
\begin{equation}
	M_{n}=\sum_{i=1}^{N} {\bf m}_i \cdot {\bf e}^{(n)}_i = \left<m|n\right> = \left<n|m\right>.
\end{equation}

In case of more than one inequivalent site exhibiting a magnetic moment, the space
of all possible magnetic configurations is spanned by
\begin{equation}
	 \left\{ \left| n_{\text{c}_1} \right> \otimes \left| n_{\text{c}_2} \right> \otimes  \dots \otimes   \left| n_{\text{c}_d} \right>  \right\},
\end{equation}
where $d$ is the number of clusters. 
Based on the above, an arbitrary magnetic configuration can be expanded as
\begin{align}
\left|m\right> =  \left| m_{\text{c}_1} \right> \otimes \left| m_{\text{c}_2} \right> \otimes  \dots \otimes   \left| m_{\text{c}_d} \right>,\\
\left|m_{\text{c}_j}\right> = \frac{1}{N^{(\text{c}_j)}}\sum^{3N^{(\text{c}_j)}}_{n=1} M^{(\text{c}_j)}_n \left|n_{\text{c}_j}\right>.
\label{eq:CMPexpansion}
\end{align}
Any two magnetic configurations on the same magnetic sites can be compared by an overlap, which we define as
\begin{equation}
\mathcal{O}_{mm'} = \left( \frac{\left<m|m'\right>}{\sqrt{\left<m|m\right>}\sqrt{\left<m'|m'\right>}}\right)^2.
\label{eq:overlap}
\end{equation}

Lastly, notice that each CMP carries a definite order and irreducible representation (irrep).
Additionally, CMPs of same order and irrep can be enumerated by a label $y$.
This is a convention to write for instance $\left| n(6 \,T_{2u};y) \right>$, where the CMP labeled $n$ is the $y$-th CMP 
of $6$th order and irrep $T_{2u}$. We recall that the $6$th order multipole is called $64$-pole in the $2^l$-nomenclature.

\subsection{Setup for SDFT} \label{Subsec:Setup for SDFT}

The \emph{ab initio} calculations are performed by the Vienna Ab initio Simulation Package (VASP) in version 5.4~\cite{Kresse1993,Kresse1994,Hobbs2001} and the flags are set appropriate to noncollinear SDFT--GGA calculation including spin-orbit coupling as described in the Supplemental Material~\cite{suppl}.
The pseudopotentials were chosen such that $d$-electrons in transition metals and $f$-electrons in Lanthanoids and Actinoids are treated as valence electrons.
The default exchange correlation functional, i.e.\ generalized gradient approximation~\cite{Perdew1996} by Perdew, Burke and Ernzerhof (PBE), is used. 

The VASP input is created by the aid of the Python Materials Genomics (pymatgen) package~\cite{Ong2013}. 
In particular, we used subroutines based on \texttt{spglib}~\cite{Togo2018}.
The magnetic configurations of the CMP basis are created by a code authored by M.-T.~Suzuki, which employs the TSPACE library~\cite{Yanase1985}.

\section{Results and Discussion} \label{Sec:Results}


In this Section, we want to explore the following two main aspects: 

(i) Is the CMP expansion a physically meaningful description of magnetic configurations? 
Namely, here the premise for a physically meaningful description constitutes that naturally occurring magnetic configurations can be characterized by one or few symmetrically related CMPs.
It can be understood in the same sense as atomic orbitals are a meaningful basis to describe electrons bound to a free atom, 
i.e.\ the probability distribution of one electron is described by one or few degenerate atomic orbitals.
In fact, this analogy extents to molecular orbitals in a complex, where the underlying spherical harmonics are symmetrized according to site symmetry.

(ii) Can SDFT predict the most stable magnetic configuration by the aid of an exhaustive list of candidate magnetic configurations for a given crystal? 
In fact, the predictive power of the combination of the CMP expansion and SDFT (CMP+SDFT) ought to be seen as a composition of the following issues:
(a) Is there evidence to assume that the list of candidate magnetic configurations generated by the CMP basis is exhaustive? 

(b) Can the experimentally determined magnetic configuration be found among all SDFT results? 
Note that the similarity between two magnetic configurations is expressed by the overlap defined in \Cref{eq:overlap}.
In addition, we compare the magnetic space group, which crucially influences physical properties.

(c) Can SDFT correctly assign the lowest total energy to the experimental magnetic configuration?

\begin{figure*}
  \centering
  \includegraphics[width=\textwidth]{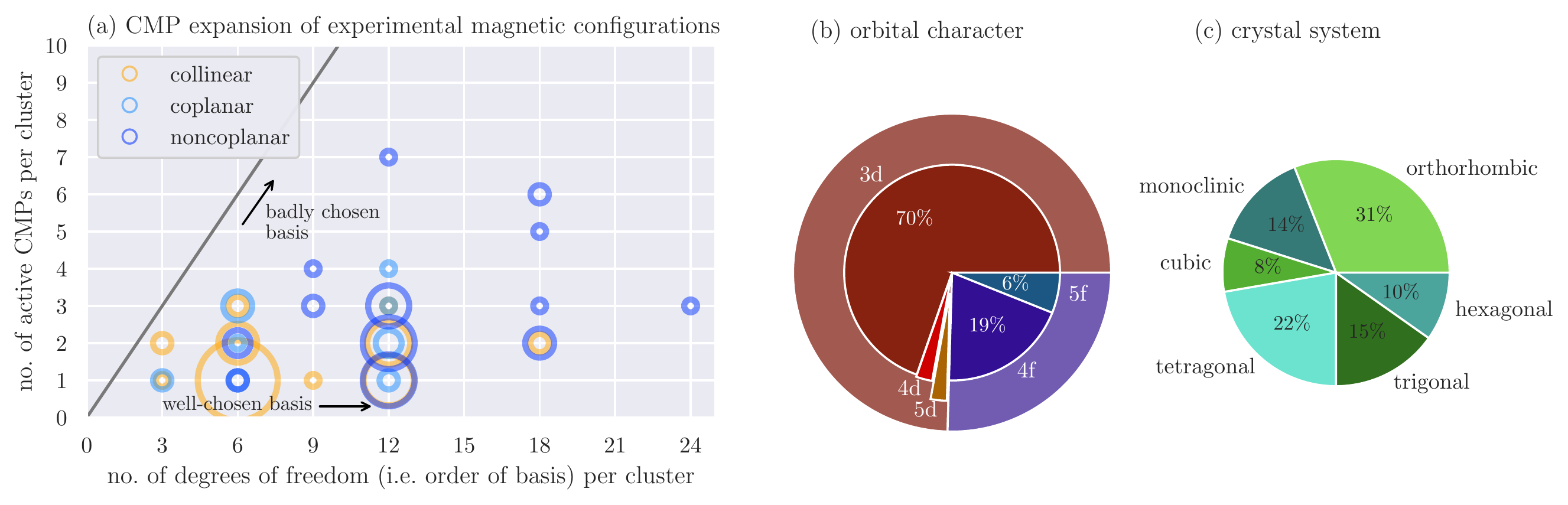}
  \caption{Properties of $\matmagorder$ experimental magnetic configurations. 
  (a) The number of CMPs needed to expand the experimental magnetic configuration---active CMPs---over the number of degrees of freedom per magnetic cluster. There are $3N$ degrees of freedom for $N$ sites in a magnetic cluster, which coincides with the order of the CMP basis. The size of the circle indicates the frequency of occurrence.
  (b) Orbital character of the magnetic site.
  (c) Crystal system. 
  \label{fig:1}
  }
\end{figure*}


\subsection{The investigated materials and workflow} 

After preluding these questions, let us start by focusing on the experimental data found on MAGNDATA~\cite{Gallego2016}. 
This commendable collection of meticulously gathered neutron diffraction measurements and other measurements, e.g.\ optomagnetic response, is
still growing and by no means complete.
The MAGNDATA entries used in this study were personally double-checked with the experimental references~\cite{Moussa1996, Tokunaga2008, White2012, Reehuis2011, Reehuis2011, Morosan2008, Kren1967, Fiebig1996, Bertaut1961, Pernet1970, Plumier1983, Brown1973, Brown1967, Alikhanov1959, Disseler2015, Mentre2008, Melot2010, Troc2012, Reynaud2013, Reynaud2013, Tsuzuki1974, Aldred1975, Melot2011, Choi2008, Hase2015, Lappas2003, Lappas2003, Schobinger-Papamantellos1997, Gitgeatpong2015, Darie2010, Schobinger-Papamantellos2012, Schobinger-Papamantellos1988, Kunnmann1968, Kunnmann1968, Avdeev2014, Battle2003, Gorbunov2016, Yamani2010, Yano2016, Taira2003, Ding2016, Redhammer2008, Yaouanc2013, Gaudet2016, Knizek2014, Scagnoli2012, Knizek2014, Ressouche2010, Blasco2016, Sanjeewa2016, Scheie2016, Ferey1985, Burlet1981, Rousse2003, Burlet1981, Wu2017, Henriques2018, Fruchart1978, Jauch2004, Singh2009, Wiedenmann1981, Zhang2015, Xu2017, Shirane1959, Garcia-Castro2018, Kim2012, Santoro1966, Steeman1990, Palacios2018, Brown1990, Burlet1997, Sukhanov2018, Cavichini2018, Hao2012, Autret2004, Brock1996, Regnault1980, Sazonov2009, Will1979, Lottermoser1988, Lottermoser1986, Wadley2015, Ouyang2005, Sale2019, Lobanov2004, Sale2017, Solana-Madruga2018, Gvozdetskyi2018, Petit2017, Hallas2017, Calder2017, Toft-Petersen2012, Calder2012, Baran2009, Schobinger-Papamantellos2001, Redhammer2009, Trump1991, Calder2012a, Fabreges2008, Blasco2017, Brown2006, Brown1981, Gukasov2002, Aczel2013, Nguyen1977, Gignoux1972, Brown2006, Rodriguez-Carvajal1991, Tomiyasu2004, Bertaut1968, Arevalo-Lopez2013, Hutanu2012, Roy2013, Munoz2000, Garlea2008, Hill2008, Munoz2000, Guo2014, Guo2014, Iikubo2008, Zhu2014, Volkova2014, Sazonov2013, Ohgushi2013, Purwanto1994, Purwanto1994, Rousse2003, Jensen2009, Calder2014, Blanco2006, Rousse2003, Brown1963, Gonzalo1966} and the specific compounds are listed in the Supplemental Material~\cite{suppl}. 

These materials explicitly contain transition metals, Lanthanoids and Actinoids with on-site magnetic moments and most data entries are fully AFM or show only weak ferromagnetism.
The magnetic configurations considered here possess zero propagation vector, which limits the available data to about $400$ entries in MAGNDATA. 
Moreover, entries corresponding to duplicates in respect to higher temperature, pressure or external magnetic field phases are excluded from this study.
Finally, some large unit cells are omitted for efficiency reasons.

The still evolving nature of this database inclined us to take a differentiated perspective on each entry:
For some materials the size of the magnetic moment is well-determined, while the magnetic order could not be uniquely identified.
And conversely, some materials have a well-known symmetry, despite the lack of an exactly determined size of the magnetic moment.
Therefore, in this study a total of $\totmat$ materials are analyzed, albeit they are distinguished in $\matmagorder$ entries with
known magnetic order and $\matmagsize$ entries with known on-site magnetic moment.

\Cref{fig:1}~(a) presents the number of CMPs needed to describe a magnetic cluster featured in the experimental magnetic configuration over the total number of degrees of freedom in the corresponding magnetic cluster. 
Here, a non-zero CMP component is a so-called \textit{active} CMP in analogy to the terminology used w.r.t.\ irreducible representations. 
The number of degrees of freedom per cluster is naturally equivalent to the order of the CMP basis. 

The data shown in~\Cref{fig:1}~(a) comprises $\expclusters$ magnetic clusters in $\matmagorder$ materials, among which $\collinear$ are classified to
be collinear, $53$ are noncollinear. In particular, $\coplanar$ are coplanar and $\noncollinear$ are noncoplanar, as indicated by the color of the circles.
Meanwhile, the size of the circle indicates the rate of occurrence.

A well-chosen basis is able to express a configuration in terms of few non-zero components.
In this regard, remarkably $\expclustersoneCMPperc\%$ of all clusters are characterized by a single active CMP. 
And only $\expclustersmorethanthreeCMP$ clusters, i.e.\ $\expclustersmorethanthreeCMPperc\%$, of the clusters in the
experimental configurations are linear combinations of more than three CMPs. 

The construction of the CMP basis~\cite{Suzuki2019} might intuitively wake the expectation that the number of active CMPs per cluster for a \textit{collinear} magnetic structure is equal or less than three.
Nevertheless, that could not have been generally expected for the \textit{noncollinear} case.
This intuition is empirically confirmed in~\Cref{fig:1}~(a), where all collinear circles are as expected reported below three active CMPs.
In the case of \textit{noncollinear} magnetic configurations, on the other hand, 
$\le 3$ contributing CMPs per cluster strongly suggests that the basis is particularly well-chosen. 
Thus, the CMP expansion of experimental configurations in~\Cref{fig:1}~(a) establishes the CMP basis to be a particularly suitable basis. 

The pie charts in~\Cref{fig:1} give an overview of the composition of all $\totmat$ materials.
In particular, \Cref{fig:1}~(b) shows the orbital character of the valence electrons on the magnetic site. 
The majority of the materials features transition metals with emerging $d$-orbital magnetism, 
while a minority of $25\%$ observes $f$-orbital magnetism.
Secondly, the pie chart in~\Cref{fig:1} (c) presents the underlying Bravais lattice and fortifies a balanced mixture comprising of all lattice types.


After we have discussed the known experimental properties, let us move on to setting up a predictive scheme.
In~\Cref{fig:2} the computational workflow is organized in four steps: input, setup, calculation, and analysis.
The input is taken in form of (magnetic) CIF files~\cite{mcif} from the database MAGNDATA. 

Step~2~in~\Cref{fig:2}, the setup, includes reading the magnetic CIF~files, 
creating the list of candidate magnetic configurations and writing the input files for VASP by the aid of pymatgen.
Crucially, in this step the CMP basis is obtained as described in~\Cref{Subsec:Cluster Multipole expansion}, which does not require
the experimental magnetic configuration as an input, but merely the choice of magnetic clusters.

We presume the following heuristic rule holds: 

\textit{The magnetic ground state favors either pure CMPs or linear combinations of CMPs that combine equally weighted CMPs of the same order and same irrep.}

This heuristic rule prompts us to extend the list of inital candidate magnetic configurations by linear combinations of same order and same irrep. 
Neglecting linear combinations of pairs yields $(Y-1)Y$ additional guesses, for $Y$ being the number of CMPs with same order and same irrep. 

In the case of more than one magnetic cluster, $d\ge2$, this would lead to too many additional guesses.
For the $\matmorethanonecluster$ materials in concern, where $d\ge2$, we chose to combine only the exact same multipole projected onto a different magnetic cluster.
In other words, the linear combination of CMPs with same order, same irrep and same $y$ is taken, c.f.~the last paragraph of \Cref{Subsec:Cluster Multipole expansion}.
Now this similarly leads to $(Y-1)Y$ additional guesses, but $Y$ is the number of CMPs, which are distinct only w.r.t.~$\text{c}_j$.

Step~3~in~\Cref{fig:2}, the VASP calculation, is performed as described in~\Cref{Subsec:Setup for SDFT}.
The total number of SDFT calculations necessary is equal to the number of candidates. 
The list of candidates is composed of in total $\left( \sum_j^d 3N^{(\text{c}_j)} \right)$ CMP basis magnetic configurations and accordingly many times $(Y-1)Y$ additional guesses.
This amounts to a total of $\totcandidates$ calculations including all $\totmat$ materials in this study. 

Step~4~in~\Cref{fig:2}, the analysis, involves determining characteristic quantities of each calculation.
First, all possible domains of the converged magnetic configuration are computed. 
To that end each space group operation combined with time reversal operations $\pm1$ is applied, which leads to
either (a) covering the magnetic configuration and thus the operation is element of the magnetic space group, or (b) a new magnetic domain.
Considering the set of operations that leave the magnetic configuration invariant, we determine the magnetic space group devising the
IDENTIFY MAGNETIC GROUP application on the Bilbao Crystallographic Server~\cite{Aroyo2011}.

All calculations of a given material and their domains are cross-checked with each other in order to filter how many distinct magnetic configurations
and, thus, distinct local minima in the SDFT total energy landscape have been identified. Quantities such as the total energy and 
the size of the magnetic moment per site are averaged over all calculation corresponding to the same local minimum.
The calculation with the lowest total energy among all SDFT calculations of a given material is the CMP+SDFT global minimum.

Note that the list of candidates created as discussed in step 3 is not free of duplicates corresponding to
different domains of the same magnetic configuration.
In the Supplemental Material~\cite{suppl} the CMP basis for YMnO$_3$ is constructed. Then, linear combinations of CMPs and magnetic domains are eluded by hands of that example.
The candidates corresponding to different domains could be excluded to avoid unnecessary numerical cost.
This amounts to a total of $\uniquecalc$ unique calculations for all $\totmat$ materials in this study,
which comprise of $\totcandidatesaddguessperc\%$ additional guesses.

To conclude step 4 in \Cref{fig:2}, all possible domains are considered when computing the overlaps of 
(i) the experimental and the initial candidate's magnetic configuration, $\mathcal{O}_{\text{exp},\text{init}}$, 
(ii) the experimental and the converged SDFT calculation's final magnetic configuration, $\mathcal{O}_{\text{exp},\text{fin}}$, 
and (iii) the initial candidate's and the converged SDFT calculation's final magnetic configuration, $\mathcal{O}_{\text{fin},\text{init}}$, as defined in \Cref{eq:overlap}.

In total this study identifies $\totlm$ CMP+SDFT local minima starting from $\uniquecalc$ unique candidates.
As mentioned, we performed $\totcandidates$ including some redundant candidates in this study.
Instead of excluding these redundant candidates that correspond to different domains of the same magnetic configuration, we used them to statistically analyze the reproducibility.
In a nutshell, the reproducibility is the probability to converge to the same local minimum, when repeating the SDFT calculation. 
More details are described in the Supplemental Material~\cite{suppl}.
In this study the reproducibility reaches $\nonuniquecandidatebutsamelm$ on a scale from $0$ to $1$, where $1$ refers to perfect reproducibility.


\begin{figure}
  \centering
\begingroup\setlength{\fboxsep}{8pt}
\colorbox{lightgray!10}{
\begin{tabular}{ p{0.9\columnwidth} }
\\
\textbf{1. Input} \\

	\begin{itemize}
		\item obtain experimental magnetic configuration from MAGNDATA as .mcif file                                                             
	\end{itemize}   	\\

\textbf{2. Setup} \\

	\begin{itemize}
		\item perform CMP expansion using Fortran code authored by M.-T. Suzuki (uses TSPACE library) 
		\item read experimental magnetic configurations and CMP basis configurations as pymatgen structure 
		\item create list of initial candidate magnetic configurations incl.\ linear combinations of same CMP order and irreducible representation
		\item write VASP input                                                             
	\end{itemize}  \\

\textbf{3. Calculation} \\

	\begin{itemize}
		\item run GGA for noncollinear magnetic magnetism in VASP  
	\end{itemize} \\

\textbf{4. Analysis} \\

	\begin{itemize}
		\item read final converged magnetic configuration as pymatgen structure
		\item determine key quantities: 
		\begin{itemize}
			\item CMP+SDFT local minima  
			\item compare total energy to obtain CMP+SDFT global minimum
			\item domains
			\item overlaps btw. initial, final, and experimental magnetic configurations
			\item magnetic space groups 
		\end{itemize}	
	\end{itemize}	 \\

\end{tabular}}\endgroup

  \caption{ Computational workflow divided in 4 steps: input, setup, calculation and analysis.
  \label{fig:2}
  }
\end{figure}


\subsection{The performance of candidate magnetic configurations} 

The high computational cost is justified, only if the list of candidates can be expected to be exhaustive. 
Let us recall that the CMP basis defined in \Cref{eq:cmp basis} spans the space of all possible magnetic configurations.
Each CMP is characterized by its order and irrep. First, we want to argue that the candidate's irrep is likely to prevail throughout the SDFT calculation. As the CMP basis is complete and, thus, any irrep that could be active in a given system explicitly appears in the CMP basis, the former corroborates that the CMP basis is a good starting point.

\begin{figure}
  \centering
  \includegraphics[width=\columnwidth]{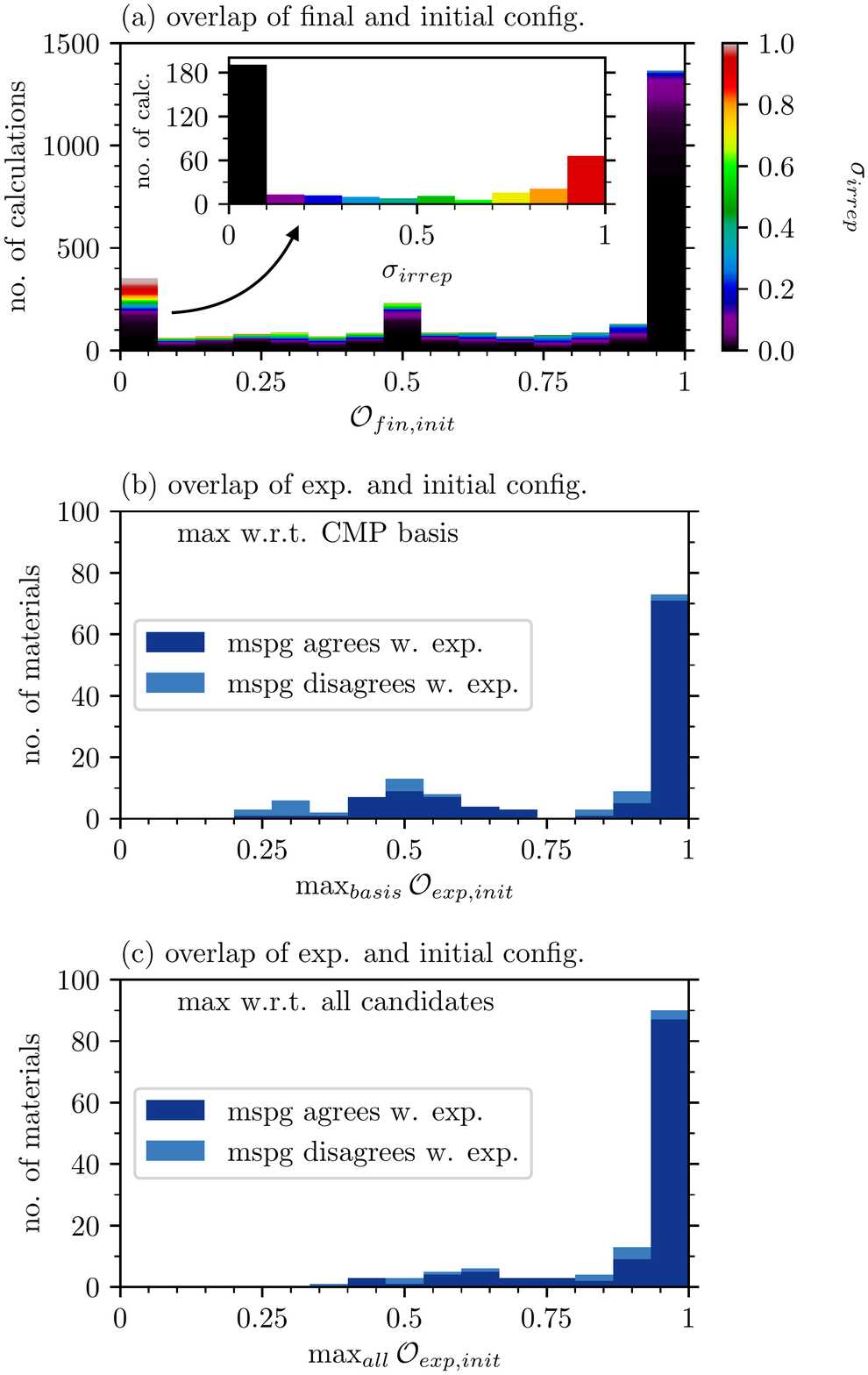}
  \caption{ (a) Overlap of the candidate and the final SDFT result.
  The overlap is defined in \Cref{eq:overlap}.
  The color scale indicates the variance of the irreducible representation. 
  Inset: The variance of the irreducible representation for the lowermost bin (see arrow). 
  (b) The maximum overlap of the experiment and the initial candidate
  w.r.t.\ the CMP basis.  
  (c) The maximum overlap of the experiment and the initial candidate
  w.r.t.\ all candidates incl.\ the CMP basis and additional guesses. 
  The color classifies if the magnetic space group (mspg) agrees with the experimentally determined mspg. 
  \label{fig:3}
  }
\end{figure}

\Cref{fig:3} (a) shows a histogram of the overlap of the final magnetic configuration and the initial candidate, $\mathcal{O}_{\text{fin},\text{init}}$.
In particular, $\mathcal{O}_{\text{fin},\text{init}} \approx 1$ corresponds to the candidate's magnetic configuration remaining almost identical during the iterations.
In that case, the candidate appears to be in close vicinity to a local minimum in the total energy landscape of SDFT. 
We see that the uppermost bin, with $\nocalcofininitinuppermostbinfigthreeaperc\%$ of all calculations, accounts for more calculations than any other bin. 

On the other hand, if the candidate does not correspond to a minimum in the total energy, 
the calculation is expected to yield a small overlap: $\mathcal{O}_{\text{fin},\text{init}} \ll 1$.
If the system converges to a magnetic configuration, which is a linear combination of the inital candidate and another magnetic configuration,
a finite $\mathcal{O}_{\text{fin},\text{init}}$ occurs.

There is a related scenario in which the system converges to a magnetic configuration that is of the same irrep, 
but does not including the CMP of the initial candidate. 
That case can be characterized by  $\mathcal{O}_{\text{fin},\text{init}} \approx 0$ and $\sigma_{\text{irrep}} = 0$, 
warranted the definition of the variance of the irrep reads
\begin{equation}
	\sigma_{\text{irrep}} =  \sum_{j,j'}^{d}  \sum_{n,n'}^{3N^{(\text{c}_j)}}   |\hat{M}^{(\text{c}_j,\text{init})}_n |
	 B_{nn'} |\hat{M}^{(\text{c}_j,\text{fin})} _{n'} |     \label{eq:var irrep}
\end{equation}
with
\begin{subequations}
\begin{align}
	\hat{M}^{(\text{c}_j,\text{init}/\text{fin})}_n &= \frac{M^{(\text{c}_j,\text{init}/\text{fin})}_n}{ \sum_{j=1}^{d} \sum_{n'=1}^{3N^{(\text{c}_j)}} \left| M^{(\text{c}_j,\text{init}/\text{fin})}_{n'} \right| },\\
	B_{nn'} &=\begin{cases}
               1,\qquad \text{irrep}_{n} \neq \text{irrep}_{n'} \\
               0, \qquad \text{irrep}_{n}  = \text{irrep}_{n'}
            \end{cases}.
\end{align}
\end{subequations}
Here, $\sigma_{\text{irrep}}$ is defined such that, if the same irreps appear with the same weight in the candidate's CMP expansion and in the
CMP expansion of the converged calculation, then $\sigma_{\text{irrep}}=0$. 
In a nutshell, $|\hat{M}_n|$ indicates to what percentage the $n$-th CMP contributes to the expansion and $B_{nn'}$ is a boolean giving zero weight to equal irreps.

The colorbar in \Cref{fig:3}~(a) corresponds to  $\sigma_{\text{irrep}}$ defined in \Cref{eq:var irrep}.
The variance of the irrep is less than $10\%$,  $\sigma_{\text{irrep}}<0.1$, in $\nocalcsigmairreplesszeropointoneperc\%$ of all SDFT calculations.
In other words, the inital irrep is highly likely to be active in the final magnetic configuration.

The inset of  \Cref{fig:3}~(a) emphasizes this observation: 
The variance of the irrep for the lowermost bin of \Cref{fig:3}~(a) is shown as a histogram.
Notably, the initial irrep has less than $10\%$ deviation, i.e.\ $\sigma_{\text{irrep}} < 0.1$, in $\figthreeainsetlowermostbinperc\%$ of the calculations with $\mathcal{O}_{\text{fin},\text{init}} \approx 0$.

As a more general statement, we have shown that the candidate's irrep is statistically likely to prevail throughout the SDFT calculation.
Conversely, the most stable magnetic configuration is less likely to be found, if the irrep is not among the list of candidates.
Hence, creating a list of candidates building upon the CMP basis is an efficient solution to assure all possible irreps are among the candidates.


From this point of view, it seems unnecessary to introduce additional guesses as candidates that are equally weighted CMPs of same order and same irrep. 
However, using the experimental data as a guide once more, the advantages of including additional guesses into the list of candidates becomes clear.

\Cref{fig:3}~(b) presents the maximum overlap of the CMP basis and the experiment, 
$\max_{\text{all}}\mathcal{O}_{\text{init},\text{exp}}$. 
The histogram shows a probability density strongly peaked close to one.
Additionally, there are side peaks at $1/3$ and $1/2$. 
This bias towards $1/3$ and $1/2$ can be appreciated when considering the aforementioned heuristic rule once again.

Namely, the magnetic ground state favors either pure CMPs or linear combinations of CMPs that combine equally weighted CMPs of the same order and same irrep.
An irrep can have a dimension of $1$, $2$ or $3$ and accordingly at each CMP order
CMPs basis configurations occur in sets of $1$, $2$ or $3$ configurations in the expansion.
Hence, favored linear combinations projected onto a CMP basis configuration are prone to yield
overlap of $1$, $1/2$ or $1/3$.

In comparison, \Cref{fig:3}~(c), displays the maximum overlap of initial candidate and the experiment, 
$\max_{\text{all}}\mathcal{O}_{\text{init},\text{exp}}$, 
w.r.t.\ the complete list of candidates, 
which contains the CMP basis configurations as well as additional guesses.
The introduction of additional guesses, following the heuristic rule, can effectively avoid side peaks at $1/3$ and $1/2$
and thus takes into account linear combinations common in materials existing in nature.

As \Cref{fig:3}~(a) showed, most magnetic configurations remain close to the initial magnetic configuration. Therefore it is paramount to start from an exhaustive list of magnetic configurations.

The dark blue and light blue colors in \Cref{fig:3}~(b) and (c) indicate, that the magnetic space group found experimentally is identical to the magnetic space group of the
candidate or not, respectively. Considering all candidates, as in \Cref{fig:3}~(c), $\nomaxexpinitmspgagrees$ of $\matmagorder$ magnetic space groups agree.
This is an improved agreement rate compared to considering only the CMP basis, as in \Cref{fig:3}~(b), where $\nomaxexpinitbasismspgagrees$ magnetic space groups agree.
It is noteworthy that some magnetic space groups only enter the list of candidates through the additional guesses.

A final argument in favor of introducing additional guesses is that in total we find $\nolmthxtoaddguess$ of $\totlm$, hence $\nolmthxtoaddguessperc\%$, of the local minima in the SDFT energy landscape only thanks to the 
additional guesses. Even among the CMP+SDFT minima with the minimum total energy $\nogmthxtoaddguess$ are thanks to the additional guesses, as well as $\noexpmostsimthxtoaddguess$ of the (local) minima most similar to the experiment.


Therefore, with the collection of arguments mentioned above, we have justified expectation that the list of candidate magnetic configurations is exhaustive.
In the following, let us investigate whether the experimentally determined magnetic configuration is present among 
all SDFT results and how we might predict the likely experimental magnetic configuration for an unknown material.

\subsection{Analysis of CMP+SDFT local minima}


\begin{figure}
  \centering
  \includegraphics[width=\columnwidth]{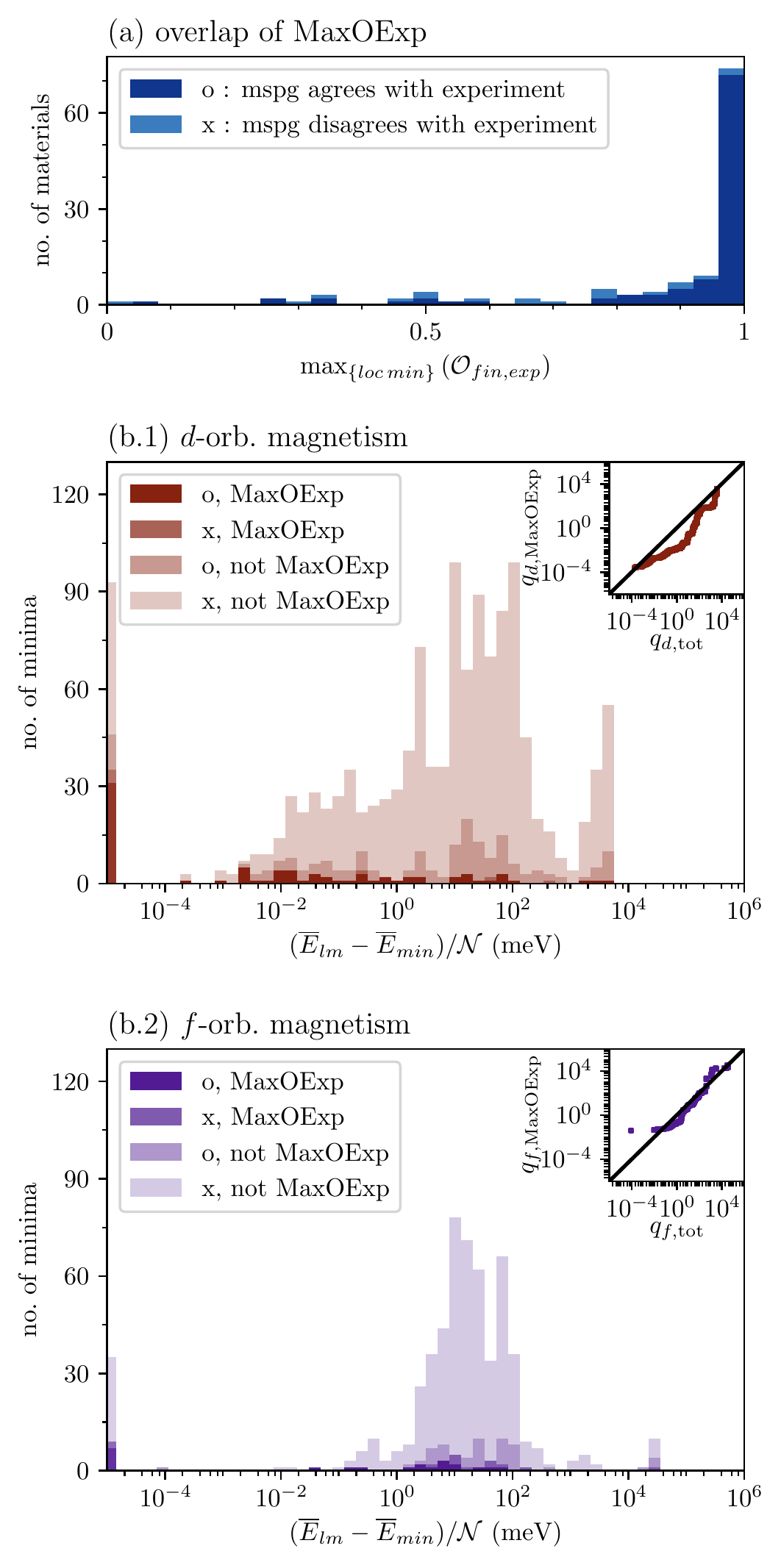}
  \caption{ (a) Overlap of experiment and CMP+SDFT minimum most similar to experiment (MaxOExp).
  Overlap is defined in \Cref{eq:overlap}.
  o/x classifies if the magnetic space group (mspg) agrees/disagrees with the experimental mspg. 
  (b.1) total energy distribution w.r.t.\ materials feature $d$-orbital magnetism.
  The minima are classified in MaxOExp 
  and remainder ``not" MaxOExp, and mspg agrees/disagrees. 
  Inset: Q-Q plot, where $q_{d,\text{MaxOExp}}$ w.r.t.\ the distribution of MaxOExp is compared to
  $q_{d,\text{tot}}$ w.r.t.\  the distribution of all local minima of materials feature $d$-orbital magnetism.
  (b.2) total energy distribution w.r.t.\ materials feature $f$-orbital magnetism.
  Inset: Q-Q plot, where $q_{f,\text{MaxOExp}}$ w.r.t.\ the distribution of MaxOExp is compared to
  $q_{f,\text{tot}}$ w.r.t.\  the distribution of all local minima of materials featuring $f$-orbital magnetism.
  \label{fig:4}
  }
\end{figure}


Following the workflow in \Cref{fig:2} all final SDFT results are scrutinized for their
similarity. Some SDFT results correspond to the same local minimum in the SDFT total energy landscape
and as such they are grouped in CMP+SDFT local minima.
The overlap of each CMP+SDFT local minimum with the experimental magnetic configuration is computed according to \Cref{eq:overlap}. 
The CMP+SDFT minimum that yields the maximum overlap with the experiment $\max_{\{loc\,min\}}{\mathcal{O}_{\text{fin},\text{exp}}}$ (MaxOExp) 
is termed to be the most similar CMP+SDFT local minimum to the experiment.  
A worthwhile run should yield $\max_{\{loc\,min\}}{\mathcal{O}_{\text{fin},\text{exp}}}\approx1$, entailing that MaxOExp is indeed very similar to the experiment.
Additionally, the magnetic space group (mspg) should agree with the experimentally detected symmetry. 

\Cref{fig:4} (a) presents $\max_{\{loc\,min\}}{\mathcal{O}_{\text{fin},\text{exp}}}$, i.e.\ the overlap of MaxOExp
for $\matmagorder$ materials with known experimental magnetic order. 
The distribution features a substantial peak at $\max_{\{loc\,min\}}{\mathcal{O}_{\text{fin},\text{exp}}}\approx1$. 
In fact, $\upperquarterfigfouraperc\%$ of MaxOExp mark $\max_{\{loc\,min\}}{\mathcal{O}_{exp,vasp}}>0.75$,
verifying good agreement of one CMP+SDFT local minimum with the experiment.

Despite the large overlap, some mspg do not agree. In particular $\upperquarterfigfouramspgagreesperc\%$ of MaxOExp agree w.r.t.\ their mspg, despite yielding $\max_{\{loc\,min\}}{\mathcal{O}_{exp,vasp}}>0.75$, see the dark blue labeled ``o". 

The upper most bin in \Cref{fig:4} (a) accumulates $\uppermostbinfigfouramspgagreesperc\%$ and corresponds to
$\max_{\{loc\,min\}}{\mathcal{O}_{exp,vasp}}>0.96$. Even in the upper most bin not all mspg agree, while on the other
hand most CMP+SDFT local minima with rather inadmissible $\max_{\{loc\,min\}}{\mathcal{O}_{exp,vasp}}<0.75$ still agree w.r.t.\ their mspg. 
For instance, Fe$_2$O$_3$ has a collinear AFM structure with a small tilting \cite{Hill2008}. 
While the parent spg is R$\overline{3}$c $(167)$ the small tilting results in P$\overline{1}$ $(2.4)$ for the mspg.
In the CPM expansion, the experimental configuration is described by two CMP basis configurations of order $5$.
However, they do not observe the same irreducible representation. 
In particular, the main contribution is A$_{1\text{g}}$ and the tilting is due to contributions of E$_{\text{u}}$.
In CMP+SDFT the most stable configuration is pure A$_{1\text{g}}$ without any tilting. 
So that, although the overlap $\max_{\{loc\,min\}}{\mathcal{O}_{exp,vasp}}=0.9658$, the mspg predicted by CMP+SDFT is R$\overline{3}$c $(167.103)$ not P$\overline{1}$ $(2.4)$ as found experimentally.

In total $\noexpmostsimmspgagreesperc\%$ among MaxOExp yield the correct mspg.
This is to say that neither the overlap nor the mspg alone are a sufficient criterion whether the experimental configuration is correctly predicted or not. 

In comparison, only $\finmspgagreesperc\%$ of all CMP+SDFT minima yield the experimental mspg.
However, for $\nomatamongalllmmspgagreesperc\%$ of the materials at least one CMP+SDFT minima yields the
experimental mspg. As mentioned among MaxOExp $\noexpmostsimmspgagreesperc\%$ yield the experimental mspg.

Another characteristic CMP+SDFT minimum is the CMP+SDFT global minimum, which observes the minimum total energy in SDFT.
Among all CMP+SDFT global minima only $\nogmmspgagreesperc\%$ yield the experimental mspg.
This shows that the mspg  of the CMP+SDFT global minima is more likely to agree with the experimental mspg than a random
CMP+SDFT minimum, but the CMP+SDFT global minima is not adequately predicting the mspg.

Let us continue by analyzing the SDFT total energy of the CMP+SDFT minima in more detail. 
Each CMP+SDFT minimum is attributed one or more SDFT results, as multiple candidates might converge
to the same minimum. An average over these attributed SDFT results leads to the material dependent and magnetic configuration
dependent total energy of a specific CMP+SDFT minimum $\overline{E}_{lm}$. 
The CMP+SDFT global minimum observes the minimum total energy $\overline{F}_{min}$.

In order to compare the total energy across materials, we take a normalized relative total energy that reads 
\begin{equation}
    (\overline{E}_{lm}-\overline{E}_{min})/\mathcal{N}.
\end{equation}
Here, $\mathcal{N}$ is the total number of degrees of freedom,
i.e.\ the sum of the order of basis over all clusters that observe a magnetic moment in SDFT in that material.
 
 \Cref{fig:4} (b.1) and (b.2) present the distribution of CMP+SDFT minima over the normalized relative total energy 
of materials featuring $d$-orbital magnetism and $f$-orbital magnetism, respectively.
The energy scale is logarithmic in units of meV. And the lowermost bin, representing the CMP+SDFT global minima,
would theoretically lie precisely at zero. However, for the obvious practical reasons, namely that $\log(0) \to -\infty$, it is added at the lower edge.
The remaining bins represent the distribution of CMP+SDFT local minima $\rho_{d/f,\text{tot}}$.
A key question is, whether MaxOExp tends to be close to the total energy minimum.

In  \Cref{fig:4} (b.1) and (b.2) the color intensity classifies all CMP+SDFT minima according to agreement/disagreement 
with the experimental mspg labeled by o/x, respectively. Additionally, the minima are classified according to being MaxOExp or not.
Overall the total energy distributions $\rho_{d/f,\text{tot}}$ span across many orders of magnitude.
Albeit, $\rho_{f,\text{tot}}$ is more concentrated in the energy range $1\meV$ up to $1000\meV$.

The data shows that in total $\nogmeqexpmostsim$ of $\matmagorder$ ($\nogmeqexpmostsimperc\%$) of the CMP+SDFT global minima 
coincide with MaxOExp. Hence, the magnetic configuration with the minimum total energy in this study
does not, at this point, identify the expected experimental configuration. 
Nevertheless, MaxOExp might tend towards smaller total energy. 
In order to gain more insight, we ask if MaxOExp data points follow the same distribution as an arbitrary local minimum in $\rho_{d/f, tot}$.

Two distributions can be compared in terms of a $\text{Q-Q}$~plot~\cite{Wilk1968}, where the $x$-axis represents the quantile of the reference distribution and the
$y$-axis represents the quantile of the sample distribution. Let us define the quantile, $q_{s/r}$ for a sample/reference distribution of local minima $\{ \text{lm}_k\}$, 
where $k=0,..,K-1$ and the local minima (lm) are ordered by $\overline{E}_{lm_k} \le \overline{E}_{lm_{k+1}}$. 
The $k/(K-1)$ quantile $q_k$ is given by
\begin{equation}
    q_k = (\overline{E}_{lm_k}-\overline{E}_{min})/\mathcal{N}.
\end{equation}
Hence, the $0.5$ quantile is simply the median value and the $0.1$ quantile is the point that divides the distribution such  that $90\%$ of the local minima have greater total energy. 

The Inset of \Cref{fig:4}~(b.1) shows the $\text{Q-Q}$~plot comparing quantiles of $\rho_{d,\text{MaxOExp}}$, as the sample distribution, 
with $\rho_{d,\text{tot}}$, as the reference distribution.
For each data point in the smaller sample distribution the quantile is computed, as explained above. 
Subsequently, $q_{d,\text{MaxOExp}}$ is juxtaposed against $q_{d,\text{tot}}$.

If the two datasets are sampled from the same underlying distribution $\rho_{d,\text{MaxOExp}}=\rho_{d,\text{tot}}$, all points align on the median. 
The quantile is defined on the same axis as the original distribution, i.e.\ $q_{d,\text{MaxOExp}}$ and $q_{d,\text{tot}}$ are defined on $(\overline{E}_{lm_k}-\overline{E}_{min})/\mathcal{N}$.

The $\text{Q-Q}$~plot in the inset of \Cref{fig:4} (b.1) shows significant deviation from the median. 
Indeed, the slow incline up to approximately $10\meV$ reveals an accumulation of MaxOExp towards lower total energy.
For $d$-orbital magnetism we find $\noexpmostsimbelowonemeVdperc\%$ of MaxOExp below $1\meV$.
On average each material has $\avglmbelowonemeVd$ CMP+SDFT local minima below $1\meV$. 
In particular, in this dataset the material with the maximum number of CMP+SDFT local minima has $\maxlmbelowonemeVd$ minima below $1\meV$.
This shows that CMP+SDFT successfully narrows down the possible magnetic configurations for a new material featuring $d$-orbital magnetism 
to a handful of CMP+SDFT local minima, that are highly likely to be close to the experimental observation.

The inset of \Cref{fig:4} (b.2) shows the analogous $\text{Q-Q}$~plot for $f$-orbital magnetism.
Here, the quantiles basically align on the median suggesting that $\rho_{f,\text{MaxOExp}}=\rho_{f,\text{tot}}$.
Moreover, for $f$-orbital magnetism we find only $\noexpmostsimbelowonemeVfperc\%$ of MaxOExp below $1\meV$.
Although in case of $f$-orbital magnetism the consideration of the total energy seems to fail in narrowing down the number of possible
magnetic configurations, at least the CMP+SDFT run itself proposes a set of $10-15$ possible magnetic configurations. 

The presented data opens a gateway to identifying a handful of magnetic configurations as CMP+SDFT local minima 
for a given material among which the experimentally stable magnetic space group and exact configuration is highly likely to be found.
Yet it has not been possible to uniquely identify the ground state based on the SDFT total energy. Although CMP+SDFT yield local minima with the experimental mspg and local minima with large overlap with the experimental magnetic configuration, SDFT fails to assign a low total energy compared to other local minima.


\begin{figure*}
  \includegraphics[width=\textwidth]{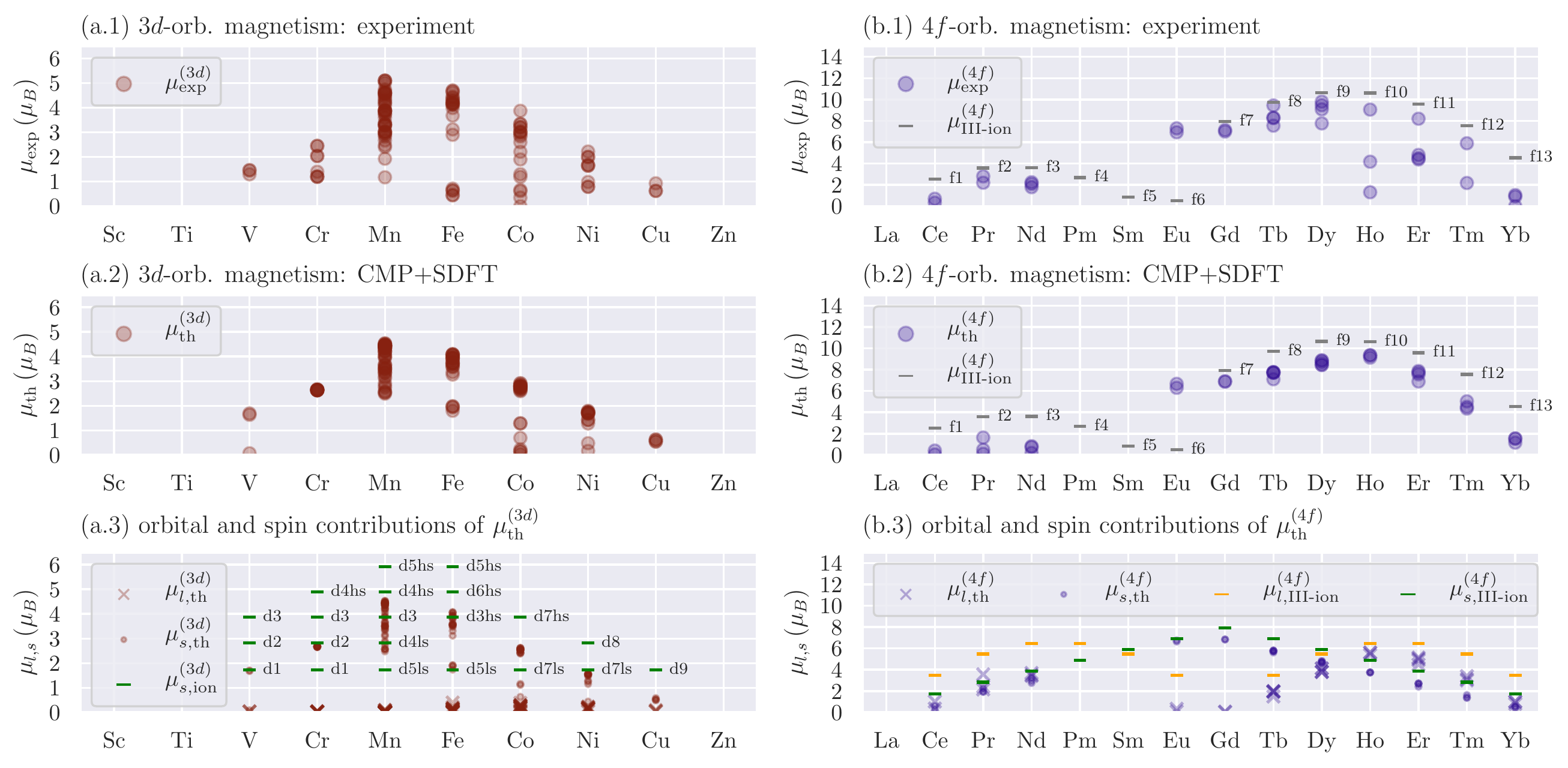}
  \caption{The magnetic moment per site as a function of electrons per atom for $3d$- and $4f$-orbital magnetism. 
  (a.1) and (b.1), the experimental magnetic moment per site $\mu_{\text{exp}}$.
  (a.2) and (b.2), the magnetic moment per site of the CMP+SDFT minimum most similar to experiment w.r.t.\ its magnetic configuration $\mu_{\text{th}}$.
  (a.3) and (b.3), absolute values of the orbital angular momentum contribution $\mu_{l}$
  and the spin contribution $\mu_{s}$ to $\mu_{\text{th}}$.
  \label{fig:5}
  }
\end{figure*}


\subsection{The magnetic moment per site}

Besides the magnetic configuration, the size of the on-site magnetic moment
crucially influences the magnetic properties of a material. Hence, it is
interesting to ask, if the magnetic moment estimated by SDFT is close to 
the experimentally determined magnetic moment per site. 
In the literature \cite{Kettle2013} it is well-known that complexes containing first row transition metals with open $3d$ orbitals are dominated by crystal field splitting. This is referred to as \textit{strong field regime}. 
Further, the ground state of complexes containing Lantanides with open $4f$ orbitals are dominated by spin-orbit coupling.
Complementary, this is referred to as \textit{weak field regime}.
Let us explore the implications by looking closer at the element-dependence of the on-site magnetic moment.

\Cref{fig:5} presents the on-site magnetic moment averaged over sites within one magnetic cluster as a function of elements sorted by increasing no.\ of electrons. 
In particular, the average magnetic moment per site reads 
\begin{equation}
    \mu_{\text{c}_j} = \frac{1}{N^{(\text{c}_j)}}  \sum_{i=1}^{N^{(\text{c}_j)}} |\vec{m}_i|
\end{equation} 
and, thus, the average is taken within each magnetic cluster $\text{c}_j$, only. The columns show the case of $3d$-orbital magnetism and $4f$-orbital magnetism, respectively.


\Cref{fig:5}~(a.1) gives an overview of the experimental results $\muexp$ for $3d$-orbital magnetism. 
We see that within compounds featuring the same magnetic element vastly different on-site magnetic moments are reported. This is referred to as \textit{compound dependence} in the following discussion.
Overall, the maximum on-site magnetic moment per element frames a dome shape with a clear maximum at Mn closely followed by Fe.
In comparison, \Cref{fig:5}~(a.2) shows the on-site magnetic moment $\muth$ predicted by CMP+SDFT.
Here, $\muth$ is taken to be the magnetic moment of the magnetic configuration with MaxOExp, which has the most similar magnetic order compared to the experiment.
We can see very good agreement in the overall tendency between experiment and CMP+SDFT.

A strong crystal field represents a real and time reversal invariant perturbation that forces a real-valued ground state which effectively quenches the orbital angular moment operator ($\vec{L}\equiv\vec{0}$) as discussed in many text books, see e.g.\ Ref.\ \cite{El-Batanouny2020}.
Therefore the spin contribution alone is expected to constitute the on-site magnetic moment.  
Fortunately, in contrast to the experiment the numeric calculation grants direct access to the spin contribution $\vec{\mu}_{s,th}$ and the angular momentum contribution $\vec{\mu}_{l,th}$ to the on-site magnetic moment
\begin{equation}
    \vec{\mu}_{th}=\vec{\mu}_{s,th}+\vec{\mu}_{l,th}.
\end{equation}

\Cref{fig:5} (a.3) presents the absolute values $\mu_{s,th}$ and $\mu_{l,th}$.
The data clearly confirms that the angular momentum is almost entirely quenched in SDFT. 
Only for the heavier elements, where spin-orbit coupling becomes more relevant \footnote{The spin-orbit coupling is proportional to $dV/dr$, where $V$ is the potential due to the ions. Hence, heavier elements exhibit stronger spin-orbit coupling. See e.g.\ Ref.\ \cite{Kuebler2017}. }, 
a small contribution is given by $\mu_{l,th}$. 
In other words, SDFT supports that for compounds with more than half-filled $3d$ bands the angular momentum is only partially quenched. 

The dominant $\mu_{s,th}$ can be directly compared to the spin-only magnetic moment in the ionic limit.
It is computed within the Russel-Saunders (or L-S) coupling scheme and is given by
\begin{equation}
    \mu_{s,\text{ion}}^{(3d)}=2\sqrt{s(s+1)}\mub
\end{equation}
with spin quantum number $s$ for the total spin operator $\vec{S}$.
The total spin $\vec{S}$ of the electronic configuration $3d^n$ with $n$ electrons is essentially constructed by following Hund's first rules. 
Albeit in real complexes the electron configuration can be in the high spin (hs) or the low spin (ls) configuration depending on the crystal field strength compared to the intra-orbital Coulomb repulsion. 
This yields different spin-only magnetic moments $\mu^{(3d)}_{s,\text{ion}}$ in the ionic limit for electronic configurations of the form $3d^n$(hs/ls).

In \Cref{fig:5} (a.3) $\mu^{(3d)}_{s,\text{ion}}$ is displayed as a reference for various possible electronic configurations. Here, we assumed octahedral complexes for the crystal field splitting. the  The maximum magnetic moment is consistent with the experiment and CMP+SDFT calculation realized for Mn$^{2+}$ or Fe$^{3+}$ in the ionic limit.
Additionally, the ionic limit already hints towards possible reasons for the observed compound dependence.
Namely, we expect the formal oxidation state and the crystal field strength to introduce compound dependence.
Further compound dependence arises due to the exact symmetry including small distortions as introduced by the Jahn-Teller effect and the choice of ligands via the nephelauxetic effect, which describes the delocalization of metal electrons through covalent bonds with the ligands.

Let us now move on to the case of compounds featuring Lanthanides shown in the right column of \Cref{fig:5}.
As mentioned, in the weak field regime spin-orbit coupling is strong compared to the crystal field effect.
Therefore the orbital angular momentum operator $\vec{L}$ cannot be neglected and the magnetic moment is computed in the $j$-$j$~coupling scheme in terms of the total angular momentum $\vec{J}$.
In the ionic limit, the electronic ground state can be determined following all three Hund's rules \footnote{For instance, the $3+$-ion for Er has $4f^{11}$ and thus $3$ unpaired spins yielding $s=3/2$. The orbital angular momentum is maximized when orbitals with magnetic quantum number $m_l=3,2,1$ are singly occupied yielding $l=6$ and L$=$I. Finally the total angular momentum $\vec{J}=\vec{S}+\vec{L}$ for more than half-filling, i.e.\ quantum number $j=15/2$. The ground state term-symbol reads $^4$I$_{15/2}$. } for a given shell configuration $4f^n$ with $n$ electrons.
The magnetic moment in terms of the total angular momentum quantum number $j$ then reads 
\begin{equation}
    \mu_j =g_j\sqrt{j(j+1)}\mub \label{eq:muion4f}
\end{equation} 
with the Land{\'e} g-factor ($g_j$). 
Representative, we compute the magnetic moment $\mu_{\text{III-ion}}^{(4f)}$ for all $3+$ ions. 
Note that in fact, Eu$^{2+}$ for instance is expected to resemble Ga$^{3+}$ because both have a $4f^7$ electronic configuration.

\Cref{fig:5}~(b.1) shows the experimental results $\muexp^{(4f)}$ for $4f$-orbital magnetism in comparison to $\mu_{\text{III-ion}}^{(4f)}$.
Similar to the $3d$-orbital magnetism, different compounds featuring the same magnetic element observe vastly different $\muexp^{(4f)}$, however the origin must be different as we will see.
A comparison to the CMP+SDFT results presented in \Cref{fig:5}~(b.2) shows good agreement of the overall characteristic behaviour.
In both, experiment and  CMP+SDFT, the magnetic moment is just below the ionic limit and a small (large) dome forms in the less (more) than half-filled region. 

\begin{figure*}
  \includegraphics[width=\textwidth]{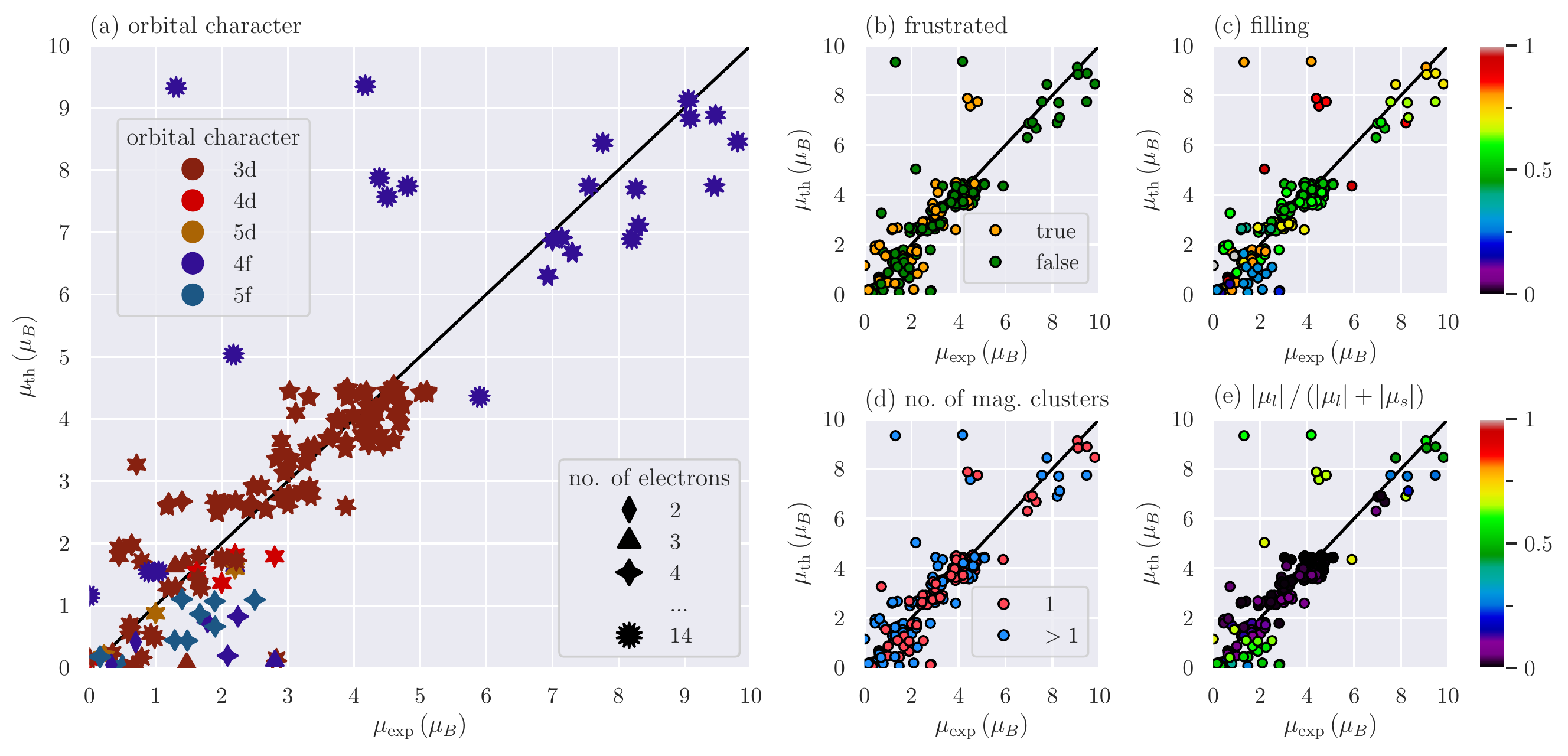}
  \caption{ Average magnetic moment per site of the CMP+SDFT minimum most similar to experiment w.r.t.\ its magnetic configuration $\mu_{th}$ compared to
  the experimentally measured magnetic moment per site $\mu_{exp}$. 
  (a) Color indicates the valence orbital character of the magnetic site. The no.\ of points on a star mark equals no.\ of electrons.
  (b) Color indicates if the material is expected to be frustrated due to odd numbered rings of antiferromagnetic bonds.
  (c) The color bar shows the filling of the valence orbital of the magnetic site.
  (d) Color indicates if a single cluster or multiple clusters are magnetic in the material. 
  (e) The color bar shows the normalized orbital angular momentum contribution $|\mu_l|/(|\mu_l|+|\mu_s|)$.
  \label{fig:6}
  }
\end{figure*}

Noticeably, the compound dependence in the CMP+SDFT results is reduced compared to the experiment.
By a more detailed analysis of the experimental data, the compound dependence in $4f$-orbital magnetism is revealed to arise when long-range order cannot be established very well experimentally.  
SDFT naturally assumes a well-established long-range order by design as it is a zero temperature method.
Specific cases are considered in the discussion of \Cref{fig:6}.

\Cref{fig:5}~(b.3) shows the absolute value of the spin and orbital contributions ($\mu^{(4f)}_{s,th}$ and $\mu^{(4f)}_{l,th}$) in SDFT. 
As a reference, we plot a fictitious spin-only $\mu_{s,\text{III-ion}}^{(4f)}$ and orbital-only magnetic moment $\mu_{l,\text{III-ion}}^{(4f)}$ in the ionic limit for $3+$ ions:
\begin{align}
    \mu_{s,\text{III-ion}}^{(4f)}=2\sqrt{s(s+1)}\mub,\\
    \mu_{l,\text{III-ion}}^{(4f)}=\sqrt{l(l+1)}\mub. 
\end{align}
Prominently, the destructively (constructively) coupling for less (more) than half-filling is confirmed and visualized.
Further, the spin contribution $\mu^{(4f)}_{s,th}$ very closely aligns with the ionic limit. 
This can be expected as $4f$ electrons barely delocalize by covalently bonding with the surrounding ligands. 
The orbital contribution $\mu^{(4f)}_{l,th}$ shows a clearly reduced value compared to $\mu_{l,\text{III-ion}}^{(4f)}$.
This might be interpreted as partial quenching of $\vec{L}$ in SDFT, 
which is supported by the observation that the reduction of $\mu_{l,\text{III-ion}}^{(4f)}$ is stronger for lighter elements. 


So far it has become clear that there is no systematic overestimation of the on-site magnetic moment by CMP+SDFT.
However naively one might anyways expect a general underestimation due to the lack of treatment
of strong electronic correlation effects in SDFT, albeit strong electronic correlation is expected in particular in $3d$ and $4f$-bands.
As we will see in the following, the data defies this general expectation of an underestimated on-site moment.
To this end, let us compare $\muth$ and $\muexp$ compound-wise, or rather cluster-wise for all compounds.


\Cref{fig:6} juxtaposes the average magnetic moment per site $\muth$ of the magnetic configuration with MaxOExp
and the experimentally measured magnetic moment per site $\muexp$. 
If for a magnetic cluster $\muth \approx \muexp$, the data point is in close vicinity to the median
and the size of the magnetic moment per site is well-estimated. 
In \Cref{fig:6} (a), each cluster $\text{c}_j$ is represented by a star, whose color indicates the orbital
character of the magnetic site and the number of points indicates which magnetic element forms the cluster.
For instance, the 5-pointed dark red star corresponds to a $\text{Mn}$-cluster, since Mn atom has five 3$d$ electrons.
At first sight, there is no general over- or underestimation seen in the scatter plot. 

Moreover, the data suggests that the uncertainty of SDFT is reflected in the absolute deviation of $|\muth-\muexp|$, 
rather than some relative deviation of the magnetic moment $|\muth-\muexp|/|\muth+\muexp|$.
Indeed, $\mupmofiveperc\%$ of the magnetic moments are within $\pm0.5\mub$, and beyond $\mupmoneperc\%$
obey $|\muth-\muexp|\le1\mub$.
Concomitantly, in the small magnetic moment regime, that is approximately $\mu \lesssim 2\mub$, 
no reliable prediction is possible. In the mid to high magnetic moment regime, on the other hand,
a mostly accurate prediction is made. 

There is an accumulation of $3d$ data points within  
$ 2\mub < \muth < 5\mub $, whose center of mass closely aligns with the median.
However, an apparent lack of precision leads to a wide spread around the median.
Despite another accumulation of $4f$ data points in the range of $6 \mub < \muth < 10\mub $ showing
similarly a high accuracy with a center of mass near the median, we also see many outliers with $4f$-orbital character
across the entire range of the on-site magnetic moments.

The specific group of three outliers at $ 4\mub < \muexp < 5\mub $ and $ 7\mub < \muth < 8\mub$ correspond to $\text{Er}$-clusters in  
$\text{Er}_2\text{Sn}_2\text{O}_7$,  $\text{Er}_2\text{Ru}_2\text{O}_7$ and $\text{Er}_2\text{Pt}_2\text{O}_7$, listed 
from left to right. 
In the ionic limit, the ground state electronic configuration of Er$^{3+}$ is $^4I_{15/2}$ with the Land{\'e} g-factor ($g_j$) of $6/5$. Therefore, $\mu^{(\text{Er})}_{\text{III-ion}}$ is estimated to be $9.58\mub$ using \Cref{eq:muion4f}. 
We see that $\muth$ of the three outliers are considerably less than $\mu_{\text{III-ion}}$.
In fact, the three outliers are known candidates for realizing a spin liquid phase due to the presence of magnetic frustration,
as described in Ref.\ \cite{Petit2017}, \cite{Taira2003} and \cite{Hallas2017}
and hence present highly non-trivial cases.

The two outliers with $\muth > 9\mub$ correspond to Ho-clusters. Both data points
are contributed by the same material $\text{HoMnO}_3$, which contains two inequivalent Ho-sites on top of a Mn-cluster. The latter orders at $T=78.5$\,K
and is well-estimated by CMP+SDFT with $\muexp^{(\text{Mn})}=3.32\mub$ and $\muth^{(\text{Mn})}=3.47\mub$.
On the other hand, experimentally ordering of the two Ho-clusters is subject to controversy \cite{Brown2006,Vajk2005,Lottermoser2004,Munoz2001,Fiebig2000}. 
It seems unclear from an experimental perspective whether one or both Ho-sites order even down to approximately $2\text{\,K}$.
Generally, the long range ordering of magnetic moments on Ho-sites is suggested 
to occur at much lower temperature compared to Mn-sites.
As mentioned above, a strict comparison of the SDFT result to $\muexp$ is inappropriate in the case that proper long-range ordering cannot be established experimentally.
Nevertheless, SDFT can be compared to the ionic limit, similar to the discussion on the three materials containing Er. The ground state electronic configuration of Ho$^{3+}$ is $^5I_{8}$, which yields $\mu^{(\text{Ho})}_{\text{III-ion}} =10.61\mub$ as an estimate. 
To conclude, in HoMnO$_3$ the  $\muth^{(\text{Ho})}$ of the Ho-clusters lie below $\mu^{(\text{Ho})}_{\text{III-ion}}$
and a strict comparison to $\muexp^{(\text{Ho})}$ is inappropriate.

In \Cref{fig:6}~(b), again $\muth$ and $\muexp$ are compared, but additionally the color indicates whether or not the compound is expected to be frustrated. 
Here, the expectation of frustration is based on whether nearest neighbors form rings of odd number of magnetic sites. 
Assuming AFM coupling
this geometrically leads to magnetic frustration.
Hence, we take advantage of the 
database being specifically focused on antiferromagnets. Furthermore, rings of even number of magnetic sites
could potentially also yield a magnetically frustrated system, if the AFM coupling is anisotropic, such as in the
Kitaev model. We hence note, that the definition of expected frustration used here is imprecise and only suitable for a quick superficial classification.

\Cref{fig:6}~(b) shows that indeed the well-estimated $4f$-clusters in the large magnetic moment regime are not expected to feature magnetic frustration. 
The discussed group of three outliers on the other hand are expected to be frustrated.
Data points with $4f$-orbital character in the small magnetic moment regime $\muth < 2\mub $ 
are likewise expected to be magnetically frustrated and are not particularly well-estimated. 
Although, we expect that $\muth$ is overestimated when the system is frustrated,
many clusters that are expected to be magnetically frustrated are not necessarily overestimated.
And some outliers are---at least in the approximate definition employed here---not expected to be frustrated.
However, as we have seen for HoMnO$_3$ there might be other non-trivial phenomena preventing a proper long-range order.
Hence, the geometrically expected magnetic frustration is not a sufficient indicator for overestimation of the magnetic moment.

\Cref{fig:6} (c) displays the filling on a colormap from 0 to 1, where 0.5 correspond to half-filling. Here, the filling is defined as the ratio between the number of $d$ or $f$ electrons in each magnetic atom and the number of orbitals. For the number of electrons, we consider the charge neutral state, i.e., the ionized state is not taken account.
Less (more) than half-filled $4f$ and $5f$-clusters appear in the underestimated (overestimated) region.

\Cref{fig:6} (d) addresses the number of magnetic clusters present in a specific compound. 
The data points corresponding to single cluster (red), and multiple clusters (blue) appear to be evenly distributed.
Let us divert the attention towards
data points with $\muth\approx 0$.
It should be noted that these are not paramagnetic solutions.
Two scenarios can yield $\muth\approx 0$: Either another cluster bears most of the on-site magnetic moment, or the spin contribution to the magnetic moment $\mu_s$ is canceled by the orbital contribution to the magnetic moment $\mu_l$.

\Cref{fig:6} (e) shows the normalized orbital contribution 
\begin{equation}
    \frac{\mu_l}{|\vec{\mu}_l|+|\vec{\mu}_s|}
\end{equation}
in SDFT to the total magnetic moment 
$\mu_{\text{th}}=|\vec{\mu}_s +\vec{\mu}_l|$.
Below the median in the small magnetic moment regime, indeed many clusters with less than half-filled orbitals
observe $\mu_l/(|\vec{\mu}_l|+|\vec{\mu}_s|) \approx 0.5$. In these instances, $\vec{\mu}_s$ and $\vec{\mu}_l$ adopt opposing signs and thus the contributions in fact cancel.
Clusters of heavier Lanthanides are well-estimated solely as a result of including $\mu_l$. 
Considering, once more \Cref{fig:5}~(b.3) and a comparison of \Cref{fig:5}~(a.1) and \Cref{fig:5}~(a.2),
the agreement between experiment and SDFT could be improved, if the orbital angular momentum would be less quenched in SDFT.

\subsection{CMP+SDFT+U case study}


\begin{figure*}
  \includegraphics[width=\textwidth]{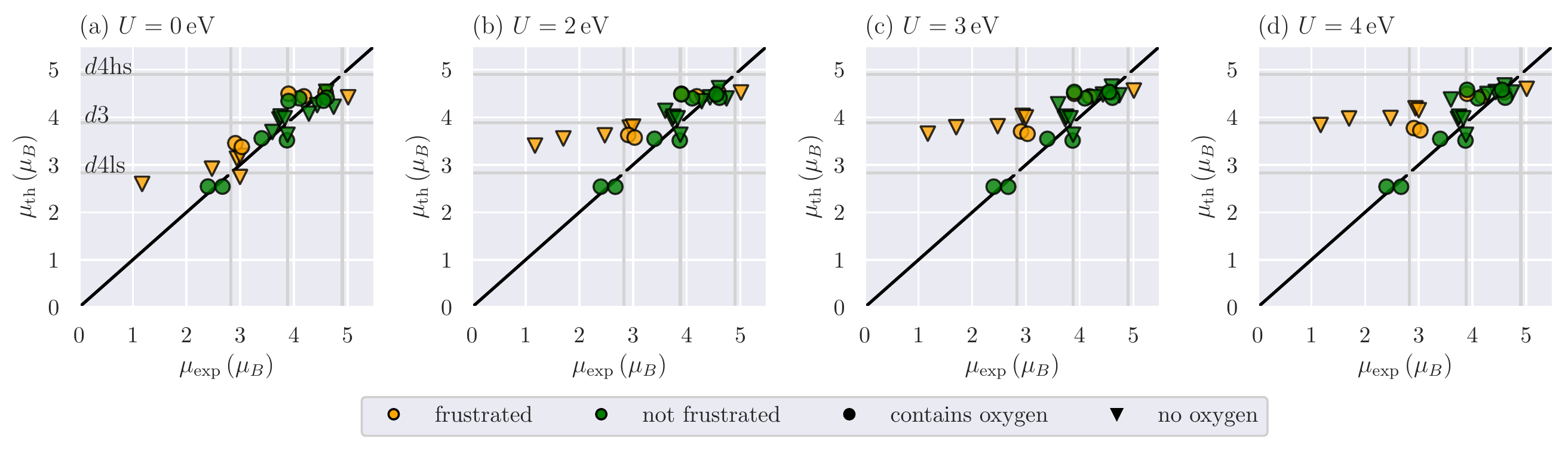}
  \caption{ Average magnetic moment per Mn-site of the CMP+SDFT+U minimum most similar to the experiment w.r.t.\ its configuration $\mu_{th}$ compared to the experimentally measured magnetic moment per site $\mu_{exp}$. (a) $U=0\,$eV, (b) $U=2\,$eV, (c) $U=3\,$eV, (d) $U=4\,$eV. 
  \label{fig:7}
  }
\end{figure*}

Hitherto we have discussed the effects of spin--orbit coupling and 
crystal field splitting on magnetism in compounds with $3d$ and $4f$-orbital character
and omitted the careful treatment of another important energy scale in these systems: the electron--electron correlation due to intra-orbital Coulomb repulsion $U$. 
There are various extensions to include electronic correlation beyond SDFT:
For instance, SDFT+U \cite{anisimov1991band}, SDFT+DMFT \cite{georges1996dynamical,kotliar2004strongly,anisimov1997first,lichtenstein1998ab,held2006realistic,held2008bandstructure}, self-consistent \emph{ab initio} D$\Gamma$A \cite{kaufmann2020selfconsistent} and other diagrammatic extensions beyond DMFT \cite{gull2011continuous,rohringer2018diagrammatic}.
In fact, these methods have brought important insight in the properties of many compounds closely related to the ones under investigation here \cite{an2011first,schlipf2013structural,kaneko2019enhanced,pulkkinen2020coulomb}, in particular w.r.t.\ Mott--Hubbard localization.

A full treatment of electron--electron correlations from first-principles for all materials introduces various challenges and is beyond the scope of this paper. 
While it is in principle possible to estimate the parameter $U$ from first-principles
by means of constraint random phase approximation (cRPA) \cite{cococcioni2005linear,csacsiouglu2011effective}, the computational cost of this procedure is immense. Therefore, albeit we aim at the prediction of the magnetic ground state from first-principles, we must resort to introducing $U$ as an adjustable parameter in this section. In particular, we will screen $U=2,3,4\,$eV for $d$-orbitals in Mn and $U=4,6,8\,$eV for $f$-orbitals of Eu and Gd in accordance with the range of typical $U$-values used in literature \cite{an2011first,pulkkinen2020coulomb}. Although this amounts to $1545$ additional CMP+SDFT+U calculations, we caution the reader, that our efforts to include $U$ may not be conclusive enough to be generalized to statements about the importance of strong electronic correlations in regard to the prediction of the magnetic ground state.

We have chosen to perform CMP+SDFT+U calculations for all materials containing a single Mn-cluster, because of the following reasons: (i) It is a well-defined subgroup of $28$ materials, which is near the minimum sample size necessary to obtain statistically significant results. (ii) The compounds are \emph{not} prone to a spin-liquid ground state, so that the comparison with the experiment stands on solid grounds. (iii) For $U=0$ the size of the magnetic moment is over- or underestimated depending on the material, as shown in \Cref{fig:7}~(a). Therefore, we can clearly distinguish if the theoretical magnetic moment $\mu_{\text{th}}$ gets closer to the experimental value $\mu_{\text{exp}}$ with increasing $U$  \emph{or} if $\mu_{\text{th}}$ increases regardless of whether it was already overestimated for $U=0$. (iv) The total energy distribution $\rho_{\text{Mn,MaxOExp}}$ of the CMP+SDFT  minimum  that  yields  the  maximum  overlap  with  the  experiment (MaxOExp) for $U=0$ has a strong bias towards the energy minimum. In other words, there is room to improve if MaxOExp were to always agreed with the CMP+SDFT global minimum and also room to deteriorate if $\rho_{\text{Mn,MaxOExp}}$ were to spread across a wider range of energy. 

Furthermore, we have chosen to perform CMP+SDFT+U calculations for all materials containing Eu and Gd, which are the following four compounds: EuTiO$_3$, EuZrO$_3$, GdVO$_4$, GdB$_4$. That is because these $4f$-elements are close to half-filling, where the orbital contribution to the on-site magnetic moment $\mu_l$ vanishes, as can be confirmed in \Cref{fig:5}~(b.3). Thus, spin--orbit coupling is of no importance in these systems and furthermore the crystal field splitting is expected to be small, because the strongly localized $4f$-orbitals are well-shielded by the outer $3d$ and $4s$-orbitals. Hence, we expect the Coulomb interaction $U$ to predominantly determine the dynamics of $f$-electrons in these compounds.

The two main questions are as follows: (i) Does including $U$ improve the prediction of the most stable magnetic structure, and (ii) will the estimation of the on-site magnetic moment improve upon introducing $U$? Without further ado let us present the results of CMP+SDFT+U for compounds containing Mn, Eu and Gd.

We find that CMP+SDFT+U identifies the same local minima as CMP+SDFT with different relative total energy to each other. Thus, MaxOExp is the same at any value of $U$. Moreover, the range of the total energy distribution $\rho_{\text{Mn,tot}}$ is $U$-independent and ranges from $0\meV$ to $1000\meV$. Hence, we ask if MaxOExp tends to have the lowest total energy and if this tendency is increased by increasing $U$. 
We note that for the limited number of materials investigated the total energy distribution of MaxOExp $\rho_{\text{Mn,MaxOExp}}$ for $U=0,2\,$eV ranges from $0\meV$ to approximately $5\meV$. Additionally the distribution of $\rho_{\text{Mn,MaxOExp}}$ is skewed towards lower total energy compared to $\rho_{\text{Mn,tot}}$. On the other hand, for $U=3,4\,$eV $\rho_{\text{Mn,MaxOExp}}$ reaches close to $1000\meV$, while it remains skewed towards lower total energy. In other words, introducing $U$ does not assign the correct total energy to the true magnetic ground state found in the experiment in this data. In fact, increasing $U$ reduces the tendency for MaxOExp to have a particularly low total energy.

Let us now discuss the estimation of the on-site magnetic moment. \Cref{fig:7}~(a) - (d) shows the average on-site magnetic moment $\muth$ of MaxOExp for all materials containing a single Mn-cluster for $U=0,2,3$ and $4$, respectively. The grey lines labeled $d4$hs (high spin), $d3$ and $d4$ls (low spin) correspond to the spin-only contribution of Mn with formal oxidation $3+$ and $4+$, i.e.\ $4$ and $3$ $d$-electrons, in an octahedral ligand-field \cite{Kettle2013}, same as in \Cref{fig:5}~(a.3). 
We see that increasing $U$ never decreases $\muth$.
For each compound we distinguish whether the crystal contains loops of odd number of magnetic sites and is thus expected to be frustrated. This is indicated by the color of the marker. The shape of the marker implies if the compound contains oxygen. Note that GGA is known to cause overbinding of oxygen to transition metals \cite{jones1989density,patton1997simplified,hammer1999improved,wang2006oxidation}. 
The effect of increasing $U$ most strongly increases $\muth$ of frustrated compounds containing no oxygen that are far away from the high spin state for $U=0\,$eV. The increase of $\muth$ also seems to occur---though less pronounced---in compounds that satisfy only one of the conditions. That is either compounds that are expected to be frustrated albeit containing oxygen or compounds lacking oxygen, although they are not expected to be frustrated.

We speculate that the overbinding of the ligand oxygen could lead to a very strong crystal field splitting. This may protect the low spin state for instance of the compounds near $\mu_{\text{th}}\approx2.5\mub$.
Furthermore, we intuitively expect frustration to reduce the size of the magnetic moment, because not all AFM bonds can be satisfied simultaneously and the cost of not satisfying a bond is proportional to the size of the on-site magnetic moment. Introducing $U$ has a localizing effect and might cause intra-atomic effects to become prevalent over frustration.
The on-site magnetic moment is reduced compared to the ionic limit due to delocalization of the Mn-electrons for instance onto the ligands. Moreover, itinerancy may lead to a reduced $\mu_{\text{th}}$ compared to the ionic limit depending on the partial density of states.
Thus, there are various reasons for the on-site magnetic moment ranging from $1\mub$ to $5\mub$. In the investigated Mn-compounds the agreement of $\muth$ with $\muexp$ corroded by introducing $U$ by means of GGA+U.

The CMP+SDFT+U results for EuTiO$_3$, EuZrO$_3$, GdVO$_4$, GdB$_4$ similarly show no improvement by introducing $U$. In fact, for EuTiO$_3$, EuZrO$_3$, GdVO$_4$ the magnetic ground state is falsely predicted to be ferromagnetic for $U>4\,$eV. Again the same local minima are found, so that in these cases MaxOExp observes increasing total energy by increasing $U$ relative to the CMP+SDFT+U global minimum at each $U$-value. For GdB$_4$ the CMP+SDFT+U global minimum is AFM along $\vec{c}$-direction for all $U$-values, while the experimental structure is a hexadecapole in the $\vec{a}\vec{b}$-plane. However, $U=0\,$eV these two magnetic structures are almost degenerate with $0.5\meV$ difference in total energy and for increasing $U$ the system increasingly prefers the out-of-plane magnetic structure. 

The on-site magnetic moment is increased with increasing $U$ for all four compounds containing Eu and Gd. As can be seen in \Cref{fig:6}~(a) around $\muexp\approx7\mub$, the size of $\muth$ is slightly underestimated for $U=0$ for all four compounds. Thus, the estimates of $\muth=6.95\mub, 6.98\mub, 7.04\mub$ and $7.10\mub$ for $U=8\,$eV for EuTiO$_3$, EuZrO$_3$, GdVO$_4$ and GdB$_4$, respectively, are closer to the experimental values $\muexp=6.93\mub, 7.30\mub, 7.00\mub$ and $7.14\mub$ than for $U=0\,$eV $\muth=6.30\mub, 6.67\mub, 6.87\mub$ and $6.91\mub$. Let us note that other $4f$-compounds observe slightly overestimates on-site magnetic moment and we suspect for these compounds increasing $U$ would also increase $\muth$.

Instead of focusing on effects of strong electronic correlations, we speculate that the prediction of the true experimental magnetic ground state could be improved by a different choice of exchange--correlation functional.
We would like to point out one recent example of a detailed SDFT+U study \cite{liu2020comparative} on LiOsO$_3$ and NaOsO$_3$ testing other exchange--correlation functionals thus far implemented in VASP, including local spin-density approximation (LSDA), PBE's improved version for solids (PBEsol), the strongly constrained appropriately normed (SCAN) meta-GGA functional and hybrid functional HSE06. By means of scanning different $U$-values including predicted ones from cRPA, Liu \emph{et al.\ }found that none of the considered functionals is capable to simultaneously predict the correct magnetic ground state for LiOsO$_3$ and NaOsO$_3$ comparing the total energy of two energetically favorable configurations. 
The treatment of exchange--correlation effects in all of these functionals hitherto implemented in VASP have the underlying assumption that locally the spin-density can be diagonalized. Schematically, an electron thus only couples to an exchange--correlation magnetic field ($\vec{B}_{xc}$) that is parallel to its own magnetization. In the last two decades, some work \cite{kleinman1999density,capelle2001spin,katsnelson2003spin,sharma2007first,scalmani2012new,bulik2013noncollinear,eich2013transverse1,eich2013transverse,pittalis2017u,goings2018current,ullrich2018density,ullrich2019spin,pluhar2019exchange} has been done to extend SDFT to include the so-called spin-torque effect, which couples the electron's spin to $\vec{B}_{xc}$ including antisymmetric terms.

\section{Conclusion and Outlook}
\label{Sec:Conclusion and Outlook}

This study is a benchmark of an \textit{ab initio} prediction of the magnetic ground state using a
novel approach termed CMP+SDFT.
This scheme devises a combination of the cluster multipole (CMP) expansion and the spin-density functional theory (SDFT) for noncollinear magnetism.
We find that materials existent in nature are well-described in terms of only few CMPs and infer the CMP basis to be a suitable basis for magnetic configurations.
Additionally, the experimental data suggests that the magnetic ground state favors either pure CMPs or linear combinations of CMPs having the same expansion order and same irreducible representation.
Guided by this heuristic rule an exhaustive list of initial candidate magnetic configurations for \emph{ab initio} calculations in the framework of SDFT is created. 

A high-throughput calculation of $\totcandidates$ \emph{ab initio} calculations using VASP led to a handful of CMP+SDFT local minima corresponding to different possible magnetic configurations for each material.
$\nomatamongalllmmspgagreesperc\%$ of materials yield the experimental magnetic space group
for at least one of the CMP+SDFT local minima. Furthermore, the maximum overlap between the experimental
magnetic configuration and the CMP+SDFT local minima exceeds
$0.75$---with $1$ corresponding to equivalence---in $\upperquarterfigfouramspgagreesperc\%$ of all materials.

An \textit{ab initio} prediction of the most stable magnetic configuration in the experiment is guided by a comparison of the total energy in SDFT using GGA of the the possible magnetic configurations for each material.
In particular, the local minimum with the larges overlap with the experiment (MaxOExp) is expected to yield the lowest total energy.
Indeed, for materials featuring magnetic sites with $d$-orbital magnetism, MaxOExp is in great majority of the cases less than $1\meV$ above the so-called CMP+SDFT global minimum. 
On the other hand, the same could not be confirmed for $f$-orbital magnetism. 
In fact, MaxOExp for $f$-orbital magnetism shows no tendency towards lower total energy. 
The implementation of GGA--PBE \cite{hobbs2000fully} used in this study did not necessarily assign the lowest total energy to the local minimum with the larges overlap with the experiment.

We have further investigated the effect of including strong electronic correlations on the level of SDFT+U for materials containing a single Mn-cluster, Eu-cluster or Gd-cluster. Our results show that for the materials we investigated introducing $U$ has a rather unfavorable influence on the prediction for both, the magnetic ground state and the size of the magnetic moment. In the end of \ref{Sec:Results}~E, we speculate that the prediction of the true experimental magnetic ground state could be improved by a different choice of exchange--correlation functional that accounts for the spin-torque effect \cite{kleinman1999density,capelle2001spin,katsnelson2003spin,sharma2007first,scalmani2012new,bulik2013noncollinear,eich2013transverse1,eich2013transverse,pittalis2017u,goings2018current,ullrich2018density,ullrich2019spin,pluhar2019exchange}, as opposed to focusing on effects of strong electronic correlations.

As far as we know, the only other scheme that aims at the prediction of noncollinear magnetic structures is based on an genetic algorithm by Zheng and Zhang \cite{zheng2020maggene}. In their approach only the fittest magnetic structures of each generation survive, which is decided based on the total energy of the magnetic structure. Thus, currently it converges to the global minimum corresponding to a theoretical magnetic ground state that is not necessarily the true magnetic ground state found in the experiment.
On the other hand, in CMP+SDFT we yield a set of magnetic configurations that are local minima of the total energy, which is very likely to include the magnetic ground state as we have demonstrated in this paper. Hence, we want to emphasize that CMP+SDFT succeeded to significantly narrow down the number of possible magnetic ground states.
This is achieved thanks to a list of candidate magnetic configurations that is tailored to account for  details of the symmetry of the crystallographic unit cell. 
In fact, CMP theory enables SDFT to identify local minima from a feasible number of candidate magnetic configurations, that put data screening and AFM material design within reach. 
On average, in this study we performed only $ 2935 / 131 = 22.4$ for each material, while in Ref.~\cite{zheng2020maggene} they performed $30$ calculations in each generation. In order to ensure convergence, they ran the evolution for $30$ generations which amounts to $900$ calculations for one material. This comparison of the number of calculations that are necessary to find the theoretical magnetic ground state, emphasizes that our list of candidates---the CMP basis combined with our heuristic rule and omitting the magnetic configurations corresponding to different magnetic domains of the same magnetic structure---is well-suited to search the space of all possible magnetic configurations.

In addition, this study showed that the on-site magnetic moment could be estimated surprisingly well by GGA without including $U$.
The precision of the predicted magnetic moment is estimated to be roughly $\pm0.5\mub$.
Some outliers arise from a lack of long-range order in the experiment.
This can be due to extremely low transition temperatures and magnetic frustration.
Despite some explainable outliers, the prediction shows no major systematic over- or underestimation of the on-site magnetic moment in GGA. 
In contrast to the experiment, the SDFT calculation grants additional insight into the balance of spin contribution and orbital angular momentum contribution to the total magnetic moment. 
The first row transition metals prove to be well-described by Russel-Saunders coupling applicable within the strong field regime. 
In other words, the orbital angular momentum is quenched and the spin-only ionic limit can be used as a reference.
The case of Lanthanides, on the other hand, is representative for systems in the weak field regime.
The on-site magnetic moment is well-described in the $j$-$j$ coupling scheme. 
In the end of \ref{Sec:Results}~D, we speculate that GGA might have slightly overestimates the crystal field effects compared to the strength of spin-orbit coupling. Some related discussions of GGA causing an overbinding of ligand oxygen can be found in the literature \cite{jones1989density,patton1997simplified,hammer1999improved,wang2006oxidation}.
This could explain why materials governed by crystal field splitting---such as the compounds with $d$-orbital magnetism---are assigned more appropriate total energy by GGA.
Yet, materials governed by spin-orbit coupling---such as Lanthanides---the experimental magnetic configuration is not assigned the lowest total energy by GGA. 
The balance between spin-orbit coupling and crystal field splitting becomes particularly crucial for lighter $4f$-elements and heavier $3d$-elements, where the orbital angular momentum is only partially quenched.


We want to end by putting this study into a bigger context and providing an outlook into future works.
The starting point of this study was the experimental database MAGNDATA~\cite{Gallego2016}.
It conveniently facilitated testing and benchmarking of our \textit{ab initio} scheme to predict the magnetic ground state.
Generally, experimental databases~\cite{Bergerhoff1983, Villars1998, White2002, Villars2010, Landolt-Boernstein, Bale2009, Linstrom2015, matweb, matnavi, Okamoto1995} 
not only facilitate testing and benchmarking of theoretical methods, but also
data mining in the experimentally explored chemical space.
Indeed, for some nonmagnetic functional materials an informed search and optimization has led to promising discoveries~\cite{Hautier2011, Berger2012, Cheng2014, Carrete2014, Zhu2015, Dunstan2016, Chen2016, Petousis2017, Dagdelen2017, Umeda2018, MansouriTehrani2018, Chen2018, Flores-Livas2019}.
However so far, apart from few pioneering works~\cite{Horton2019,sanvito2017accelerated,stevanovic2012correcting,gorai2016thermoelectricity,Xu2020} that are constrained to specific cases,
these breakthroughs in material design have not yet been matched by similar advances with respect to AFM materials.
Certainly one of the major obstacles is that compared to databases of crystal structures with more than $200\,000$ entries, MAGNDATA has to date a modest amount of about $1\,130$ entries. 
This is because the experimental determination of the magnetic configuration is much more involved than that of the crystal structure. 
Given this situation, it is an urgent challenge to construct a large-scale computational database of AFM materials. The presented benchmark provides a crucial step in laying a solid foundation for the construction of such a computational database of AFM materials.
We are optimistic that \emph{ab initio} calculations will soon be able to reliably predict the magnetic ground state. Based on that, our CMP+SDFT scheme will be able to construct a computational database of magnetic materials with a feasible amount of computational effort. 
On top of that database, model calculations---using for instance the Liechtenstein method \cite{liechtenstein1987local,udvardi2003first,turek2006exchange}---can lead to useful insights in particular w.r.t.\ the spin wave dispersion and critical temperatures of magnetic phase transitions. Finally, let us note that many magnetic transitions are accompanied by structural transitions. And it might prove imperative to follow a scheme of successively relaxing the atomic position and the magnetic ground state. In the current study we avoided this obstacle by using the atomic positions obtained experimentally. However, in view of material design, the ability to treat experimentally unknown crystal structures will be of great use.

\section{Acknowledgments}

We are thankful for useful discussions with Stephan Huebsch, Takashi Koretsune and Saeed Bahramy.

Moreover, we gratefully acknowledge the Center for Computational Materials Science, Institute for Materials Research, Tohoku University for the use of MASAMUNE-IMR (MAterials science Supercomputing system for Advanced MUlti-scale simulations towards NExt-generation - Institute for Materials Research). (Project No.\ 19S0005)

This work was supported by a Grant-in-Aid for Scientific Research (No.\ 19H05825, and No.\ 16H06345) from the Ministry of Education, Culture, Sports, Science and Technology, and  CREST (JPMJCR18T3) from the Japan Science and Technology Agency, and JSPS  KAKENHI Grants Numbers JP15H05883 (J-Physics), JP19H01842, JP20H05262 and JP20K21067, and JST PRESTO Grant number JPMJPR17N8.

\bibliography{benchlib}

\end{document}

%% file: data.tex
\newcommand{\totmat}{131}

\newcommand{\matmagorder}{122}
\newcommand{\expclusters}{162}

\newcommand{\matmagsize}{116}

\newcommand{\expclustersoneCMPperc}{48.77}

\newcommand{\expclustersmorethanthreeCMP}{6}
\newcommand{\expclustersmorethanthreeCMPperc}{3.70}

\newcommand{\matmorethanonecluster}{73}

\newcommand{\totcandidates}{2935}

\newcommand{\totcandidatesaddguessperc}{35.75}

\newcommand{\totlm}{2005}

\newcommand{\nocalcofininitinuppermostbinfigthreeaperc}{46.54}

\newcommand{\nocalcsigmairreplesszeropointoneperc}{78.16}

\newcommand{\uniquecalc}{2313}
\newcommand{\nonuniquecandidatebutsamelm}{0.79}

\newcommand{\figthreeainsetlowermostbinperc}{53.95}

\newcommand{\nomaxexpinitmspgagrees}{117}
\newcommand{\nomaxexpinitbasismspgagrees}{110}

\newcommand{\nolmthxtoaddguess}{655}
\newcommand{\nolmthxtoaddguessperc}{32.67}
\newcommand{\nogmthxtoaddguess}{23}
\newcommand{\noexpmostsimthxtoaddguess}{15}

\newcommand{\upperquarterfigfouraperc}{82.44}

\newcommand{\uppermostbinfigfouramspgagreesperc}{54.96}

\newcommand{\upperquarterfigfouramspgagreesperc}{70.99}

\newcommand{\noexpmostsimmspgagreesperc}{84.43}

\newcommand{\nogmmspgagreesperc}{37.70}

\newcommand{\nomatamongalllmmspgagreesperc}{90.16}

\newcommand{\nogmeqexpmostsim}{43}
\newcommand{\nogmeqexpmostsimperc}{35.25}

\newcommand{\finmspgagreesperc}{16.17}

\newcommand{\noexpmostsimbelowonemeVdperc}{77.66}
\newcommand{\avglmbelowonemeVd}{4.45}
\newcommand{\maxlmbelowonemeVd}{18}

\newcommand{\noexpmostsimbelowonemeVfperc}{32.43}

\newcommand{\collinear}{69}
\newcommand{\noncollinear}{43}
\newcommand{\coplanar}{10}

\newcommand{\mupmofiveperc}{51.90}

\newcommand{\mupmoneperc}{77.22}